\documentclass[pdflatex,iicol,sn-mathphys-num]{sn-jnl}

\usepackage{amsmath,amssymb}
\usepackage{graphicx}
\renewcommand{\orcidlogo}{%
  \raisebox{-0.15ex}{\includegraphics[height=1.35ex]{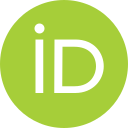}}}
\usepackage{capt-of}
\usepackage{booktabs}

\newlength{\epjcfootnoterulespace}
\let\epjcoriginalfootnoterule\footnoterule
\renewcommand{\footnoterule}{%
  \kern-\epjcfootnoterulespace\epjcoriginalfootnoterule}

\newcommand{\objectwidthfraction}{0.975}
\renewcommand{\tablecaptionfont}{\normalfont\fontsize{8bp}{9.5bp}\selectfont}
\renewcommand{\figurecaptionfont}{\normalfont\fontsize{8bp}{9.5bp}\selectfont}
\renewcommand{\tablebodyfont}{\normalfont\fontsize{9.5bp}{11.5bp}\selectfont}
\newenvironment{inlinetable}{%
  \par\addvspace{10pt}\noindent\hfill\begin{minipage}{\objectwidthfraction\columnwidth}%
  \centering\tablebodyfont
}{%
  \end{minipage}\hfill\null\par\addvspace{10pt}%
}
\newenvironment{inlinefigure}{%
  \par\addvspace{10pt}\noindent\hfill\begin{minipage}{\objectwidthfraction\columnwidth}%
  \centering
}{%
  \end{minipage}\hfill\null\par\addvspace{10pt}%
}
\renewcommand{\Authorfont}{\fontsize{11bp}{13bp}\selectfont\titraggedcenter}
\renewcommand{\addressfont}{\fontsize{9bp}{11bp}\selectfont\titraggedcenter}
\usepackage{etoolbox}
\makeatletter
\patchcmd{\@maketitle}{\vskip24pt}{\vskip12pt}{}{\PackageError{manuscript}{Title contact spacing patch failed}{}}
\patchcmd{\@maketitle}{\vskip24pt}{\vskip12pt}{}{\PackageError{manuscript}{Title abstract spacing patch failed}{}}
\patchcmd{\@maketitle}{\vskip36pt}{\vskip24pt}{}{\PackageError{manuscript}{Title bottom spacing patch failed}{}}
\makeatother
\hypersetup{colorlinks=true,linkcolor=blue,citecolor=blue,urlcolor=blue,filecolor=blue}

\begin{document}

\title[Impact of errors-on-errors on the PDF determination]{Impact of errors-on-errors on the PDF determination from combined HERA data}

\author*[1]{\fnm{Enzo} \sur{Canonero}\,\orcid{https://orcid.org/0000-0002-7180-4562}}\email{\href{mailto:Enzo.Canonero@rhul.ac.uk}{Enzo.Canonero@rhul.ac.uk}}
\author*[1]{\fnm{Glen} \sur{Cowan}\,\orcid{https://orcid.org/0000-0001-8363-9827}}\email{\href{mailto:G.Cowan@rhul.ac.uk}{G.Cowan@rhul.ac.uk}}
\author[1]{\fnm{Georgia} \sur{Brown}\,\orcid{https://orcid.org/0009-0004-2376-1336}}
\makeatletter
\g@addto@macro\artauthors{%
  \par and the xFitter developers:\par
  \global\punctcount=7\relax\def\authorsep{}}
\makeatother
\author[2,3]{\fnm{Hamed} \sur{Abdolmaleki}\,\orcid{https://orcid.org/0000-0002-2017-7706}}
\author[4,5]{\fnm{Francesco} \sur{Giuli}\,\orcid{https://orcid.org/0000-0002-8506-274X}}
\author[6]{\fnm{Alexander} \sur{Glazov}\,\orcid{https://orcid.org/0000-0002-8553-7338}}
\author[7]{\fnm{Francesco} \sur{Hautmann}\,\orcid{https://orcid.org/0000-0001-7563-687X}}
\author[8]{\fnm{Fredrick} \sur{Olness}\,\orcid{https://orcid.org/0000-0001-6799-2436}}
\author[9]{\fnm{Pavel} \sur{Starovoitov}\,\orcid{https://orcid.org/0000-0003-1990-0992}}
\author[10]{\fnm{Oleksandr} \sur{Zenaiev}\,\orcid{https://orcid.org/0000-0003-3783-6330}}

\affil[1]{Physics Department, Royal Holloway, University of London,
Egham TW20 0EX, United Kingdom}
\affil[2]{Department of Physics, Malayer University, Malayer, Iran}
\affil[3]{School of Physics, Institute for Research in Fundamental Sciences (IPM),
Tehran, Iran}
\affil[4]{INFN, Sezione di Roma Tor Vergata,
Via della Ricerca Scientifica 1, 00133 Rome, Italy}
\affil[5]{Universit\`a degli Studi Link,
Via del Casale di S. Pio V, 44, 00165 Rome, Italy}
\affil[6]{Deutsches Elektronen-Synchrotron (DESY),
Notkestrasse 85, 22607 Hamburg, Germany}
\affil[7]{Physics Department, University of Oxford, Oxford, United Kingdom}
\affil[8]{Department of Physics, Southern Methodist University,
Dallas, TX 75275-0175, USA}
\affil[9]{University of Sharjah, 27272 Sharjah, UAE}
\affil[10]{II. Institute for Theoretical Physics, University of Hamburg,
Luruper Chaussee 149, 22761 Hamburg, Germany}

\abstract{We study how the determination of parton distribution functions (PDFs) from the
combined HERA inclusive deep-inelastic-scattering data is affected when the
quoted sizes of the systematic errors are themselves treated as
uncertain.  We refer to these uncertainties informally as errors-on-errors and account for them using the Gamma Variance Model, which treats the squared quoted systematic errors as Gamma-distributed estimates of unknown variances. As a result, the usual quadratic nuisance parameter constraints
in the $\chi^2$ are replaced with logarithmic ones.  We present two applications of the framework.  In the first, a robustness test
estimates the largest errors-on-errors effects that could be present in our reproduction of the HERAPDF2.0 NNLO fit, reducing
$\chi^2/n_{\rm dof}$ from $1363/1131$ to $1306/1131$,
with the main impact concentrated at low Bjorken $x$ and low squared momentum
transfer, $Q^2$.  Comparable reductions are observed with a more flexible
small-$x$ parametrisation, both at fixed NNLO and with NNLO+NLL$_x$
resummation.  In the second, the error-on-error parameters are instead assigned a priori, as they would be in a PDF
determination intended for publication.  Taken together, the results indicate that errors-on-errors can account for a
substantial part of the apparent data--theory tension in the combined HERA
dataset, improving the inferred compatibility and widening the fitted PDF
uncertainties without appreciably displacing the PDFs or cross-section
predictions themselves.}

\maketitle
\raggedbottom

\section{Introduction}
\label{sec:intro}

Modern global parton distribution function (PDF) determinations obtain their
constraining power by combining precision measurements from different
experiments, processes and kinematic regions. This variety of inputs is
essential for separating parton flavours and determining their dependence on
$x$, the proton momentum fraction carried by the parton,
but can also expose tensions when datasets that are individually well described
prefer different PDFs in a simultaneous fit.  Such tensions are often reflected in poor goodness of fit, with values of
$\chi^2/n_{\rm dof}$ (the reduced $\chi^2$) significantly above unity,
where $\chi^2$ is evaluated at its minimum and $n_{\rm dof}$ denotes
the number of degrees of freedom.

The combined HERA I+II inclusive deep-inelastic-scattering (DIS)
dataset~\cite{H1:2015ubc}, released with its reference QCD analysis \mbox{HERAPDF2.0},
is a cornerstone of modern global PDF determinations. It brings together the H1
and ZEUS neutral- and charged-current $e^{\pm}p$ cross sections over a broad
range in Bjorken $x$ and $Q^2$, the negative squared four-momentum transfer.
The lower-$Q^2$ neutral-current measurements constrain
the low-$x$ sea and, through their scale dependence, the gluon, while the
high-$Q^2$ neutral- and charged-current data constrain the valence
distributions. At next-to-next-to-leading order (NNLO), the \mbox{HERAPDF2.0} fit to
these data yields
$\chi^2/n_{\rm dof}=1363/1131$, well above the range expected from
nominal statistical fluctuations. 

Several effects may contribute to an increase in the observed $\chi^2$, including
limitations of the perturbative calculation or heavy-flavour treatment,
insufficient flexibility of the input PDF parametrisation and assumptions in the experimental uncertainty and
correlation model. For example, Bonvini and Giuli~\cite{Bonvini:2019wxf} showed that a more
flexible low-$x$ PDF parametrisation reduced the minimum $\chi^2$ by 62 units compared with the fit using
the \mbox{HERAPDF2.0} parametrisation.
They also found that including next-to-leading logarithmic small-$x$
(NLL$_x$) resummation matched to NNLO reduced the minimum $\chi^2$
by 28--29 units relative to the corresponding fixed-order NNLO fit.
This comparison used the separate APFEL+HELL/\mbox{FONLL-C} setup introduced
in Ref.~\cite{Abdolmaleki:2018jln}.

We focus here on imperfect knowledge of the quoted systematic
uncertainties of the input measurements. Their magnitudes are often estimated from calibration or simulated
samples of finite size, or by varying modelling assumptions and
analysis choices. Limited sample sizes and assumptions about the choice and range
of these variations introduce uncertainty into the estimated
systematic-error scales themselves. We refer to this additional uncertainty as \emph{errors-on-errors}. Treating the quoted systematic uncertainties as exact can make discrepancies between data and theory appear more significant than they are, particularly when those uncertainties are underestimated. This can bias the extracted PDFs and lead to underestimated uncertainties. We relax this approximation by explicitly accounting for uncertainty in the systematic-error scales within the fit using the Gamma
Variance Model (GVM), introduced in~\cite{Cowan:2018gqq} and further
developed in~\cite{Canonero:2023qiz,Canonero:2024zlb}.

Previously, the broader issue of imperfect experimental-uncertainty modelling
has been investigated from several perspectives: PDF fits to differential
$t\bar t$ data have shown a marked sensitivity to the assumed correlation model
of the dominant systematic uncertainties, with a non-negligible effect on the
fitted gluon~\cite{Bailey:2019yze}; the effect of deliberately introduced data
inconsistencies on the fitted PDFs has been studied with idealised
pseudodata~\cite{Watt:2012tq}; and closure tests have been used to assess
whether PDF uncertainties remain reliable when correlated experimental
uncertainties are underestimated~\cite{Barontini:2025}. In these approaches, the quoted uncertainties enter the fit as exact inputs; here instead the uncertainty in the systematic-error scales is introduced explicitly as part of the statistical model through the GVM.

We apply this framework to the combined HERA inclusive dataset by implementing
the GVM in \textsc{xFitter}, an open-source framework for QCD analyses and PDF
determinations~\mbox{\cite{Alekhin:2014irh,Bertone:2017tig,xFitter:2022zjb}}.
Sections~\ref{sec:modifiedchi2} and~\ref{sec:herafitsetup} present the model,
implementation and fit setup. We then consider two complementary applications. First, Sec.~\ref{sec:robustness} presents a \emph{robustness test} designed to
estimate the largest plausible impact of errors-on-errors on the minimum
$\chi^2$, fitted PDFs and cross-section predictions. To distinguish flexibility in the uncertainty model from limitations
of the PDF shape and small-$x$ theory, Sec.~\ref{sec:interplay} repeats
the study with the more flexible Bonvini--Giuli parametrisation and
with NNLO+NLL$_x$ resummed predictions. Second, Sec.~\ref{sec:informed} presents an illustrative
\emph{informed application}, in which errors-on-errors are assigned a priori
using information external to the PDF fit rather than tuned to reduce
$\chi^2$. Conclusions are given in Sec.~\ref{sec:conclusions}.

\section{Errors-on-errors in PDF fits}
\label{sec:modifiedchi2}

The GVM provides a statistical treatment of
uncertainties in the assigned sizes of systematic errors, referred to here as
errors-on-errors~\cite{Cowan:2018gqq,Canonero:2023qiz,Canonero:2024zlb}.  We denote the conditional density of the measurements by
\(P(\boldsymbol{y}\mid\boldsymbol{a},\boldsymbol{b})\), where
\(\boldsymbol{y}\) is the vector of measured values,
\(\boldsymbol{a}\) denotes the parameters of interest (POIs) entering
the theoretical predictions, and \(\boldsymbol{b}\) denotes the nuisance
parameters (NPs) introduced to account for systematic uncertainties.
For PDF fits, \(\boldsymbol{a}\) denotes the parameters defining the PDFs
at the input scale, while systematic effects are modelled as biases
in the measured values \(\boldsymbol{y}\) that depend linearly
on \(\boldsymbol{b}\).
To follow the notation commonly used in the PDF-fitting community, these
parameters are denoted by \(\boldsymbol{a}\) and \(\boldsymbol{b}\), rather
than by \(\boldsymbol{\mu}\) and \(\boldsymbol{\theta}\), as in
Refs.~\cite{Cowan:2018gqq,Canonero:2023qiz,Canonero:2024zlb}.

For each systematic source $s$, the NP $b_s$ is constrained by a Gaussian auxiliary estimate
$u_s$, which represents the best estimate of the bias induced in the data by the source.
The uncertainty assigned to this estimate, $\sigma_{u_s}$, is commonly referred to
as the systematic uncertainty for source $s$.  The GVM treats the square of this uncertainty, i.e. the variance $\sigma_{u_s}^2$,
as an adjustable parameter and models its estimate $v_s$ as Gamma-distributed. This choice is
motivated by the sample variance of independent observations from a Gaussian
distribution, which follows a scaled $\chi^2$ distribution and hence a Gamma
distribution; see Ref.~\cite{Cowan:2018gqq} for further details. Up to terms
independent of the fitted parameters, the full $\chi^2$ function of the model is
\begin{equation}
 \begin{aligned}
 &\chi^2(\boldsymbol{a},\boldsymbol{b},
                  \boldsymbol{\sigma}_{\boldsymbol{u}}^2)
 =-2\log P(\boldsymbol{y}\mid\boldsymbol{a},\boldsymbol{b})
 \\
 &\quad+\sum_s\Biggl[
 \frac{(u_s-b_s)^2}{\sigma_{u_s}^2}
 +\left(1+\frac{1}{2\epsilon_s^2}\right)
  \log\sigma_{u_s}^2
 \\
 &\hspace{4em}+\frac{v_s}{2\epsilon_s^2\sigma_{u_s}^2}
 \Biggr].
 \end{aligned}
 \label{eq:gvm_full}
\end{equation}
The parameter $\epsilon_s$ quantifies how well the systematic uncertainty
$\sigma_{u_s}$ is known: it represents the relative uncertainty on its estimate
$\sqrt{v_s}$, and is therefore referred to as the error-on-error parameter.\footnote{More precisely,
$\epsilon_s$ is half the relative uncertainty on $v_s$.
By linear error propagation, it approximates the relative uncertainty on
$\sqrt{v_s}$, with good accuracy for $\epsilon_s\lesssim1$;
see App.~A of Ref.~\cite{Cowan:2018gqq}.}
For example, $\epsilon_s=0.3$ corresponds approximately to a $30\%$
relative uncertainty on the quoted size of the systematic error.
We adopt a frequentist approach and profile over the systematic
variances at fixed $\boldsymbol{a}$ and $\boldsymbol{b}$.\footnote{
Profiling means minimising $\chi^2$ with respect to a subset of
parameters for each fixed choice of the remaining ones. The resulting
profiled $\chi^2$ depends only on the remaining parameters.}
This gives
\begin{equation}
 \widehat{\sigma}_{u_s}^2(b_s)=
 \frac{v_s+2\epsilon_s^2(u_s-b_s)^2}
      {1+2\epsilon_s^2}.
 \label{eq:gvm_profiled_variance}
\end{equation}
Substituting into Eq.~(\ref{eq:gvm_full}) and subtracting additive
terms independent of the fitted parameters gives
\begin{equation}
 \begin{aligned}
 &\chi^2(\boldsymbol{a},\boldsymbol{b}) =
 -2\log P(\boldsymbol{y}\mid\boldsymbol{a},\boldsymbol{b})
 \\
 &{}+\sum_s\left(1+\frac{1}{2\epsilon_s^2}\right)
 \log\left[1+2\epsilon_s^2
 \frac{(u_s-b_s)^2}{v_s}\right].
 \end{aligned}
 \label{eq:gvm_profiled}
\end{equation}
This expression shows how uncertainty in the size of a systematic error
affects the constraint on the corresponding NP. A larger
$\epsilon_s$ means that the nominal estimate of $\sigma_{u_s}$ is less precisely
known. The logarithmic constraint accounts for this by allowing greater freedom
for large shifts of the NP away from its auxiliary estimate $u_s$: the penalty
for such shifts grows more slowly than in the usual quadratic treatment. The
fit can therefore accommodate experimental tensions associated with that source
through a larger systematic adjustment. For small $\epsilon_s$, expanding the logarithmic terms in Eq.~(\ref{eq:gvm_profiled}) gives
\begin{equation}
 \begin{aligned}
 &\left(1+\frac{1}{2\epsilon_s^2}\right)
 \log\left[1+2\epsilon_s^2\frac{(u_s-b_s)^2}{v_s}\right]
 \\
 &\qquad=\frac{(u_s-b_s)^2}{v_s}+\mathcal{O}(\epsilon_s^2),
 \end{aligned}
 \label{eq:gvm_gaussian_limit}
\end{equation}
so the usual quadratic constraint is recovered as $\epsilon_s\rightarrow0$.

The best estimates of the model parameters, denoted by
$\widehat{\boldsymbol{a}}$ and $\widehat{\boldsymbol{b}}$, are defined by
\begin{equation}
 (\widehat{\boldsymbol{a}},\widehat{\boldsymbol{b}})
 =\underset{\boldsymbol{a},\boldsymbol{b}}{\operatorname{arg\,min}}\,
 \chi^2(\boldsymbol{a},\boldsymbol{b}).
 \label{eq:mle_definition}
\end{equation}
These are the maximum-likelihood estimators.
The minimum $\chi^2_{\rm min}=\chi^2(\widehat{\boldsymbol{a}},
\widehat{\boldsymbol{b}})$ is used as the goodness-of-fit statistic.  It
measures how well the fitted model describes the data relative to their
assigned uncertainties.  In the Gaussian limit and for a correctly specified
model it is approximately $\chi^2$-distributed with
$n_{\rm dof}=N_{\rm data}-n_a$ degrees of freedom,
where $N_{\rm data}$ is the number of cross-section
measurements and $n_a$ the number of
fitted POIs. Hence, for a satisfactory fit, the reduced $\chi^2$ should be close to unity,
with relative fluctuations of order $\sqrt{2/n_{\rm dof}}$. Significantly larger values indicate a poor description of the data or
underestimated uncertainties, whereas significantly smaller values can
indicate overestimated uncertainties or overfitting.

Confidence intervals in parameter space can be constructed from the $\chi^2$ function in two
closely related ways. In the $\Delta\chi^2$ approach, a value of a given
parameter is retained as long as the increase from the global minimum,
$\Delta\chi^2=\chi^2-\chi^2_{\min}$, stays below a prescribed threshold. Alternatively, near the minimum the $\chi^2$ surface can be approximated quadratically, with
its local curvatures and parameter correlations encoded in the Hessian
\begin{equation}
 H\equiv\left.\frac{1}{2}
 \nabla_{(\boldsymbol{a},\boldsymbol{b})}^2\chi^2
 \right|_{(\widehat{\boldsymbol{a}},\widehat{\boldsymbol{b}})}.
 \label{eq:hessian_definition}
\end{equation}
In this quadratic limit $H^{-1}$ is the estimated covariance matrix of the
fitted parameters and the two constructions coincide: for a single parameter
of interest, the standard choice $\Delta\chi^2\leq1$ gives an approximate
$68.3\%$ confidence interval,
$\widehat a_m\pm\sqrt{(H^{-1})_{a_ma_m}}$. Away from the quadratic regime the
two differ, the $\Delta\chi^2$ intervals being in general asymmetric. For
further details on the Hessian method for computing uncertainties in PDF fits,
see, e.g., Ref.~\cite{Pumplin:2001ct}.
For nonzero errors-on-errors, corrections are needed when using the
minimum $\chi^2$ to assess goodness of fit and when constructing
confidence intervals, as discussed in Sec.~\ref{sec:profiling_bartlett}.

\subsection{Specialisation to \texorpdfstring{\textnormal{\textsc{xFitter}}}{xFitter}}
\label{sec:xfitter_chi2}

To apply the GVM to PDF fits, we implement it in
\textsc{xFitter}~\mbox{\cite{Alekhin:2014irh,Bertone:2017tig,xFitter:2022zjb}}, an open-source
framework for QCD analyses and PDF determinations used by both the theory and
experimental communities. The \mbox{HERAPDF2.0} analysis was performed within this
framework, then known as \textsc{HERAFitter}, and \textsc{xFitter} has since
been widely used in LHC measurements to assess their impact on PDFs; see, for
example,
Refs.~\cite{ATLAS:2016WZPDF,ATLAS:2021VJetsPDF,ATLAS:2022PDF21,
CMS:2019WCharm,CMS:2022JetsPDF}.

For the present study, we use the \mbox{HERAPDF2.0} $\chi^2$ definition implemented
in \textsc{xFitter}. At fixed values of the NPs, the likelihood is a product
of independent Gaussian terms for the individual measurements.
The statistical and uncorrelated systematic uncertainties determine
the point-by-point variance, while correlated systematic effects enter
through NP shifts.\footnote{Uncorrelated systematic uncertainties are
treated as independent between measurements. A correlated systematic
source affects several measurements through a shared NP.} The
resulting data contribution to the $\chi^2$ is
\begin{equation}
 \begin{aligned}
 &-2\log P(\boldsymbol{y}\mid\boldsymbol{a},\boldsymbol{b})
 \\
 &=\sum_i
 \frac{\left[y_i-T_i(\boldsymbol{a})
 \left(1-\sum_s\gamma_{is}b_s\right)\right]^2}
 {\delta_{i,{\rm stat}}^2y_iT_i(\boldsymbol{a})+
  \delta_{i,{\rm uncor}}^2T_i^2(\boldsymbol{a})}
 \\
 &{}+\sum_i\log\left(
 \frac{\delta_{i,{\rm uncor}}^2T_i^2(\boldsymbol{a})+
       \delta_{i,{\rm stat}}^2y_iT_i(\boldsymbol{a})}
      {(\delta_{i,{\rm uncor}}^2+
        \delta_{i,{\rm stat}}^2)y_i^2}\right).
 \end{aligned}
 \label{eq:xfitter_probability}
\end{equation}
Here, $y_i$ denotes the measured cross section and $T_i(\boldsymbol{a})$ the
corresponding theoretical prediction.  The coefficients $\gamma_{is}$ describe the relative response of
measurement $i$ to systematic source $s$. Since the same NP controls
the effect of that source on all affected measurements, it links
their shifts and introduces correlations. The quantities
$\delta_{i,{\rm stat}}$ and $\delta_{i,{\rm uncor}}$ denote the relative
statistical and uncorrelated systematic uncertainties. The uncorrelated systematic variance scales with $T_i^2$, corresponding
to a multiplicative uncertainty. The statistical term instead uses
$y_iT_i$, reflecting Poisson counting statistics.  The second sum is the
logarithmic normalisation required because this variance depends on the PDF
parameters $\boldsymbol{a}$ through $T_i(\boldsymbol{a})$; its parameter-independent denominator makes the logarithmic term
vanish when $T_i=y_i$, without changing the fitted parameter values. Further details on the construction of this $\chi^2$ can be found in Ref.~\cite{H1:2015ubc}.

To incorporate errors-on-errors, we substitute the data contribution
in Eq.~(\ref{eq:xfitter_probability}) into the profiled GVM of
Eq.~(\ref{eq:gvm_profiled}). We assume that any known systematic bias
has already been removed from the data, so that $u_s=0$.
We express $b_s$ in units of the nominal systematic uncertainty,
absorbing this scale into the corresponding column of the
$\gamma_{is}$ matrix, so that $v_s=1$. With these conventions,
the resulting $\chi^2$ is
\begin{equation}
 \begin{aligned}
 \chi^2(\boldsymbol{a}, \boldsymbol{b}) ={}& -2\log
 P(\boldsymbol{y}\mid\boldsymbol{a},\boldsymbol{b})
 +\sum_p b_p^2
 \\
 &+\sum_q\left(1+\frac{1}{2\epsilon_q^2}\right)
 \log\left(1+2\epsilon_q^2b_q^2\right).
 \end{aligned}
 \label{eq:eoe_xfitter_chi2}
\end{equation}
The index $p$ runs over correlated systematic sources whose uncertainties are
treated as exactly known, whereas $q$ runs over correlated systematic sources
assigned a non-zero error-on-error $\epsilon_q$.
The use of relative systematic errors does not alter the GVM construction:
for each source, $b_s$ rescales the complete absolute response vector with
components $T_i\gamma_{is}$.

In this application, errors-on-errors affect the correlated systematic
sources, while the statistical and aggregated uncorrelated uncertainties in
Eq.~(\ref{eq:xfitter_probability}) are unchanged. The contributions of individual sources to the uncorrelated systematic
uncertainty could also be treated through NPs, as discussed in
Ref.~\cite{Canonero:2024zlb}. However, the combined HERA dataset provides only the total uncorrelated
systematic uncertainty for each measurement, without a breakdown by source.

\subsection{Departures from the Gaussian limit}
\label{sec:profiling_bartlett}

For nonzero error-on-error parameters, the logarithmic NP constraints in
Eq.~(\ref{eq:eoe_xfitter_chi2}) depart from the quadratic form recovered in
the Gaussian limit of Eq.~(\ref{eq:gvm_gaussian_limit}). This departure has two practical consequences. The NP profiling equations,
used to minimise $\chi^2$ over the NPs at fixed PDF parameters, are linear
for quadratic constraints but become nonlinear with errors-on-errors.
Moreover, the minimum $\chi^2$ can depart from its usual approximate
distribution, and the confidence intervals described
above may be too narrow,
requiring statistical corrections. We summarise the treatment of both effects
here, while their derivations and implementation details are given in
Apps.~\ref{app:profiling} and~\ref{app:bartlett}.

The PDF parameters $\boldsymbol{a}$ are determined numerically.  The NPs can
either be included in this numerical minimisation or profiled for each
hypothesised value of $\boldsymbol{a}$.  The former, referred to here as
external minimisation, provides the reference solution of
Eq.~(\ref{eq:eoe_xfitter_chi2}) but increases the dimension of the numerical
problem by the number of correlated systematic sources.  For quadratic constraints,
the NPs can be profiled analytically in closed form, avoiding this increase in
the dimension of the numerical minimisation; see, e.g.,
Ref.~\cite{Stump:2001gu}.  Its
extension to the GVM is, however, non-trivial, since the logarithmic
constraints lead to coupled nonlinear equations for the NPs.  The iterative
procedure summarised below provides an approximate analytic profiling of this
system. This method was developed in Ref.~\cite{Canonero:2024zlb}, where the
perturbative results were checked against numerical minimisation.

To write the profiling equations in a compact form, we define
\begin{equation}
 \begin{aligned}
 r_i(\boldsymbol{a})&=y_i-T_i(\boldsymbol{a}),\\
 \sigma_i^2(\boldsymbol{a})&=
 \delta_{i,{\rm stat}}^2y_iT_i(\boldsymbol{a})+
 \delta_{i,{\rm uncor}}^2T_i^2(\boldsymbol{a}),\\
 \Gamma_{is}(\boldsymbol{a})&=-T_i(\boldsymbol{a})\gamma_{is}.
 \end{aligned}
 \label{eq:profiling_definitions}
\end{equation}
For fixed $\boldsymbol{a}$, the NPs are updated at iteration $k$ according to
\begin{equation}
 b_s^{(k+1)}
 =
 \sum_p\left(A^{(k)}\right)^{-1}_{sp}B_p.
 \label{eq:iterative_profiling_summary}
\end{equation}
Here, the matrix $A^{(k)}$ and vector $B$ are defined as
\begin{equation}
 \begin{aligned}
 A_{sp}^{(k)}&=
 \frac{\delta_{sp}}{\hat\sigma^{2\,(k)}_{u_s}}
 +\sum_i\frac{\Gamma_{is}\Gamma_{ip}}{\sigma_i^2},\\
 B_s&=\sum_i\frac{\Gamma_{is}r_i}{\sigma_i^2},
 \end{aligned}
\end{equation}
where the systematic variance is re-evaluated at each iteration from Eq.~(\ref{eq:gvm_profiled_variance}),
\begin{equation}
 \hat\sigma^{2\,(k)}_{u_s}
 =
 \frac{1+2\epsilon_s^2\left(b_s^{(k)}\right)^2}
      {1+2\epsilon_s^2}.
\end{equation}
The procedure is initialised with $\hat\sigma^{2\,(0)}_{u_s}=1$, so the first
iteration reproduces the analytic Gaussian profiling solution for quadratic NP
constraints.  Subsequent iterations update the systematic variances and thus account for the
non-quadratic GVM terms.  At convergence, the solution minimises the $\chi^2$
with respect to the NPs at fixed $\boldsymbol{a}$.  Sources with large NP shifts or
poor convergence are instead minimised numerically.  The derivation, convergence
condition and numerical validation for the HERA application are given in
App.~\ref{app:profiling}.

With the profiling treatment in place, we now turn to the consequences of
the logarithmic constraints for the goodness-of-fit statistic and confidence
intervals. The goodness-of-fit statistic $\chi^2_{\rm min}$, obtained by
minimising Eq.~(\ref{eq:eoe_xfitter_chi2}), is not exactly
$\chi^2$-distributed, because the logarithmic NP constraints
depart from the quadratic Gaussian form underlying the usual $\chi^2$
approximation. Furthermore, confidence intervals based on a quadratic
approximation near the minimum may be too narrow. Ignoring these effects slightly
underestimates dataset compatibility and yields narrower PDF uncertainties.
To correct for these effects, we introduce two correction factors, denoted by
$C_{\chi^2}$ and $C_{\rm CI}$, following the rescaling method introduced by
Bartlett~\cite{Bartlett:1937} and subsequently developed in
Refs.~\cite{Cordeiro:1993,CordeiroCribariNeto:2014}. Its application to the GVM
was studied in Refs.~\cite{Canonero:2023qiz,Canonero:2024zlb}, including
validation with Monte Carlo simulations in the models considered.

$C_{\chi^2}$ rescales the goodness-of-fit statistic, while
$C_{\rm CI}$ rescales the squared confidence-interval (CI) widths. These factors are defined as
\begin{equation}
 \begin{aligned}
 C_{\chi^2}&=1+\frac{b_{\chi^2}}{n_{\rm dof}},\\
 C_{\rm CI}&=1+\frac{b_{\rm CI}}{n_a},
 \end{aligned}
 \label{eq:bartlett_correction_coefficients}
\end{equation}
where $n_{\rm dof}=N_{\rm data}-n_a$ is the number of degrees of freedom,
$N_{\rm data}$ is the number of fitted measurements, and $n_a$ is the number
of fitted PDF parameters (POIs). The raw minimum is divided by $C_{\chi^2}$
and each interval half-width is multiplied by $\sqrt{C_{\rm CI}}$, without
changing the best-fit parameter values:
\begin{equation}
 \begin{aligned}
 \chi^2_{\rm B}&=\frac{\chi^2_{\rm min}}{C_{\chi^2}},\\
 \Delta a_{m,\pm}^{\rm B}&=\sqrt{C_{\rm CI}}\,\Delta a_{m,\pm}.
 \end{aligned}
 \label{eq:bartlett_applied_factors}
\end{equation}
Here $\Delta a_{m,-}$ and $\Delta a_{m,+}$ denote the lower and upper
half-widths of the uncorrected confidence interval for parameter $a_m$,
respectively.  The leading correction terms $b_{\chi^2}$ and $b_{\rm CI}$ are obtained from the analytical expressions
developed by Lawley~\cite{Lawley:1956} and Cordeiro~\cite{Cordeiro:1993} and specialised to the GVM
in Refs.~\cite{Canonero:2023qiz,Canonero:2024zlb}. Here we express the Bartlett correction in matrix form for direct
use in \textsc{xFitter}; details are given in App.~\ref{app:bartlett}.

\subsection{Relation to tolerance prescriptions}

A common method used in global PDF fits to address experimental tensions
and obtain more realistic confidence intervals is to introduce a tolerance.
In its simplest form, this prescription enlarges the nominal one-parameter
confidence region from $\Delta\chi^2\leq1$ to $\Delta\chi^2\leq T^2$,
using a fixed tolerance $T>1$. More refined prescriptions allow $T$
to depend on the direction in PDF-parameter
space~\cite{Watt:2012tq,MSHT20}.  
Ideally one would prefer to avoid introducing an ad hoc tolerance, as it complicates the statistical interpretation of the resulting parameter region.

Tolerance prescriptions enlarge the confidence region without modifying
the objective function (the $\chi^2$ minimised in the fit) or attributing
the enlargement to individual experimental uncertainty sources.
The GVM instead modifies the objective function by allowing for uncertainty
in the sizes of systematic errors whose sources are assigned nonzero
$\epsilon_s$. Within the assumed GVM model, improvements in
data--theory compatibility can therefore be traced to specific sources,
helping to identify uncertainty estimates that warrant experimental
reassessment.

The two approaches can nevertheless be complementary.
The study by M. Reader~\cite{Reader2024} of ATLAS $W$, $Z$ and
inclusive-jet data within the MSHT framework suggests that errors-on-errors
can account for part of an effective tolerance, with
$T^2\sim1.5$--$2$ in the examples considered. The tolerance required in
PDF fits may also reflect other effects, such as uncertainties in the
assumed bin-to-bin correlations of individual systematic sources.
In a combined approach, the tolerance would be applied when constructing
confidence intervals from the GVM objective function. Since this function
already accounts for the assigned errors-on-errors, the tolerance would
address the remaining tensions. One would therefore expect a smaller
tolerance than in an otherwise equivalent fit with Gaussian constraints.

\section{\mbox{HERAPDF2.0}, NNLO, RTOPT fit setup}
\label{sec:herafitsetup}

The fits are performed using the combined HERA I+II inclusive reduced cross sections measured by the
H1 and ZEUS experiments~\cite{H1:2015ubc}.
The combined dataset comprises seven samples: four neutral-current (NC)
$e^+p$ samples at proton-beam energies of 920, 820, 575 and 460~GeV;
one NC $e^-p$ sample; and two charged-current (CC) samples for
$e^+p$ and $e^-p$ scattering. The latter three are quoted at a
reference proton-beam energy of 920~GeV. The DIS kinematics are described by $Q^2$, the negative squared
four-momentum transfer, and Bjorken $x$, the dimensionless scaling
variable. Following the \mbox{HERAPDF2.0} analysis, we retain measurements
with $Q^2\geq3.5~\mathrm{GeV}^2$. This leaves 1145 measurements,
covering $x$ values from about $4\times10^{-5}$ to $0.65$ and extending
up to $Q^2=5\times10^4~\mathrm{GeV}^2$. In broad terms, the low-$Q^2$ NC data
constrain the low-$x$ sea and, through their scale dependence, the gluon,
whereas the difference between the high-$Q^2$ NC $e^+p$ and $e^-p$ cross
sections, together with the CC data, constrains the valence distributions.
The complementary flavour sensitivities of the NC and CC measurements help
separate the up- and down-type quark combinations.

The predictions are calculated at NNLO in QCD in the $\overline{\mathrm{MS}}$
scheme.  The PDFs are evolved with
\textsc{Qcdnum}~\cite{Botje:2010ay}, using renormalisation and factorisation
scales $\mu_R^2=\mu_F^2=Q^2$.  Following the \mbox{HERAPDF2.0} setup, the heavy-quark
coefficient functions entering the NC structure functions are evaluated in
the optimised Thorne--Roberts (RTOPT) general-mass variable-flavour-number
scheme, whereas the zero-mass approximation is used for the CC structure
functions, since all fitted HERA CC data satisfy $Q^2\gg M_b^2$%
~\cite{Thorne:1997ga,Thorne:2006qt,Thorne:2012az}.
The evolution uses a variable-flavour-number scheme with $n_f^{\max}=5$,
heavy-quark mass parameters $M_c=1.43~\mathrm{GeV}$ and
$M_b=4.50~\mathrm{GeV}$, and fixed $\alpha_s(M_Z)=0.118$.  Electroweak effects
are included at leading order, using the boson masses
$M_Z=91.1876~\mathrm{GeV}$ and $M_W=80.385~\mathrm{GeV}$, the weak mixing
angle $\theta_W$ specified by $\sin^2\theta_W=0.23127$, and the Fermi constant
$G_F=1.16638\times10^{-5}~\mathrm{GeV}^{-2}$.

At the starting scale $Q_0^2=1.9~\mathrm{GeV}^2$, below the charm threshold,
five distributions are parametrised: $g$, $u_v$, $d_v$,
$\bar U=\bar u$ and $\bar D=\bar d+\bar s$.
Here $g$ is the gluon distribution, $u_v=u-\bar u$ and $d_v=d-\bar d$ are the
up- and down-valence distributions, and $\bar U$ and $\bar D$ are the up- and
down-type sea-antiquark combinations, respectively.  Apart from the
additional gluon term specified below, their generic form is
\begin{equation}
 xf(x,Q_0^2)=A_f x^{B_f}(1-x)^{C_f}P_f(x),
 \label{eq:hera_pdf_parametrisation}
\end{equation}
where $A_f$ is an overall normalisation coefficient and $B_f$ and $C_f$
control the low- and high-$x$ behaviour.  The factor $P_f(x)$ is a polynomial
that allows additional shape flexibility where required.  For $u_v$ and
$\bar U$, it is chosen as $P_{u_v}(x)=1+E_{u_v}x^2$ and
$P_{\bar U}(x)=1+D_{\bar U}x$, respectively, while $P_f(x)=1$ for $d_v$,
$\bar D$ and the leading gluon term.  The full gluon parametrisation contains
an additional negative term, $-A'_g x^{B'_g}(1-x)^{C'_g}$, with a separate
fitted low-$x$ exponent $B'_g$ and $C'_g=25$.  The coefficients $A_g$,
$A_{u_v}$ and $A_{d_v}$ are fixed by the momentum and quark-number sum rules.
At $Q_0^2$, the remaining constraints are $B_{\bar U}=B_{\bar D}$,
$s=\bar s=f_s\bar D$ with the strange suppression factor $f_s=0.4$, and
$A_{\bar U}=(1-f_s)A_{\bar D}$.  Together, these give the standard
14-parameter \mbox{HERAPDF2.0} fit and hence
$n_{\rm dof}=1145-14=1131$.

For each combined measurement, the HERA combination~\cite{H1:2015ubc} provides a
statistical uncertainty, one aggregate uncorrelated systematic uncertainty and
169 correlated systematic errors. The uncorrelated systematic components
were added in quadrature before averaging, so their source-by-source breakdown
is not available. The correlated sources, by contrast, are resolved individually.  Of the 169, 162 originate from the H1 and ZEUS measurements, including
normalisations; they were propagated through the joint fit of the combined cross
sections and published in an orthogonal basis, and are denoted
\texttt{sysHZComb}.  The remaining seven are
procedural, introduced by the combination itself: they cover the choice between
additive and multiplicative errors, possible correlations of the photoproduction
background and hadronic energy scale across datasets, and four modes associated
with groups of sources having unusually large pulls.  Setting $\epsilon_s=0$ for all 169 correlated sources, this setup gives
$\chi^2_{\rm min}=1363$ for 1131 degrees of freedom, in agreement with the
published \mbox{HERAPDF2.0} NNLO result, $\chi^2_{\rm min}/n_{\rm dof}=1363/1131$.
To distinguish this setup from those introduced later, we label it
\emph{\mbox{HERAPDF2.0}, NNLO, RTOPT} and refer to the fit with Gaussian
constraints as the corresponding Gaussian baseline.

\section{Robustness study on HERA data}
\label{sec:robustness}

The first application we present uses the errors-on-errors framework as a robustness test of the \mbox{HERAPDF2.0}, NNLO, RTOPT fit. Rather than assigning the uncertainty on each systematic source, we ask what the largest realistic effect of errors-on-errors on the analysis is. In particular, we assess how much of the observed tension they could account for, how they would affect the fitted PDFs and cross-section predictions, and in which kinematic regions these effects would be concentrated. The values of $\epsilon_s$ are therefore chosen to probe the largest plausible effect, rather than being assigned a priori from source-specific knowledge. This type of study is especially useful when such information is unavailable,
as in retrospective analyses where the original expert knowledge is no longer
accessible. It can also assess the potential impact of errors-on-errors and
identify sources that merit experimental follow-up, without committing to
precise $\epsilon_s$ values for individual sources. An a priori assignment of $\epsilon_s$ is considered in Sec.~\ref{sec:informed}.

\subsection{Forward selection}
\label{sec:hera_forward_selection}

To identify the correlated systematic sources to which the fit is most
sensitive, we build the set of selected sources iteratively. Starting from the
\mbox{HERAPDF2.0}, NNLO, RTOPT baseline, in which all $\epsilon_s=0$, we first test each of the 169
correlated sources in turn by assigning that source $\epsilon_s=0.6$ and
refitting the PDF parameters and NPs. The source producing the largest reduction in the Bartlett-corrected
minimum $\chi^2_{\rm B}$, defined in
Eq.~(\ref{eq:bartlett_applied_factors}), is retained.
We use $\chi^2_{\rm B}$ because it provides a more accurate measure
of data--theory compatibility when errors-on-errors are nonzero. At
the next step, this source remains assigned $\epsilon_s=0.6$, and each of the
remaining sources is tested one at a time in addition to it. The source giving
the largest further reduction is again retained. The same procedure is then
repeated: after each step, all previously selected sources remain active, while
each remaining source is tested in turn on top of the updated fit. This
iterative procedure is referred to as \emph{forward selection}. At each step, the improvement from adding a candidate source $s$ is quantified
by
\begin{equation}
 \begin{aligned}
 \delta\chi^2_{\rm B}(s)={}&
 \chi^2_{\rm B}(\text{current fit})\\
 &-\chi^2_{\rm B}(\text{current fit}+s).
 \end{aligned}
 \label{eq:dchi2_criterion}
\end{equation}
The iteration stops when the best remaining source gives
$\delta\chi^2_{\rm B}\leq1$. A simpler scan in which each source is tested only
against the initial baseline can provide a useful first picture at lower
computational cost, but its reductions cannot be combined directly because
adding one source changes the fit minimum and therefore the impact of the
others. We set $\epsilon_s=0.6$ to probe a large but plausible effect.
As a reference, $\epsilon=1/\sqrt{2}\simeq0.7$ corresponds to the value one would assign to a
``two-point systematic'', for which the uncertainty size is estimated from
the spread of two independent determinations.\footnote{For a sample variance
based on $n$ independent observations from the same Gaussian distribution,
\mbox{$\epsilon=1/\sqrt{2(n-1)}$}. Thus $n=2$ gives $\epsilon=1/\sqrt{2}$;
see Ref.~\cite{Cowan:2018gqq}.}
Our reference value lies slightly below this benchmark; however, the choice of $\epsilon_s$ is indicative: one could also consider values
such as $0.5$ or $0.7$, or explore the dependence through a scan, as done
below. The emphasis is on understanding the impact for large but plausible
values of $\epsilon_s$, rather than on fine-tuning a parameter whose value
can itself only be estimated approximately.

The selection reported in Table~\ref{tab:forward} identifies seven correlated
systematic sources and lowers $\chi^2_{\rm B}$ by $57$ units. The leading
source, \texttt{sysHZComb1053}, accounts for $37$ units, while the remaining
six contribute about $21$ units. In an ongoing analysis or an active experiment,
the large effect associated with the leading source would motivate experimental
follow-up to reassess its uncertainty estimate and its impact on the measurement.

A preliminary indication of sources worth testing for errors-on-errors can be obtained from sizeable NP shifts in the Gaussian baseline fit, expressed in units of the nominal systematic uncertainty. The magnitude of a shift alone, however, does not determine the effect, since the improvement also depends on the data constraints and correlations with the other fitted parameters. Further details on NP shifts and their
uncertainties are given in App.~\ref{app:hera_np_pulls}.

\begin{inlinetable}
\centering
\captionof{table}{Forward selection for \mbox{HERAPDF2.0}, NNLO, RTOPT, with $\epsilon_s=0.6$.
Rows 1--7 list the selected sources in order.
$\chi^2_{\rm B}$ is the corrected minimum after each addition,
and $\delta\chi^2_{\rm B}$ its reduction from the previous step.}
\label{tab:forward}
\tablebodyfont
\renewcommand{\arraystretch}{1.25}
\setlength{\tabcolsep}{2.5pt}
\begin{tabular*}{\linewidth}{@{\extracolsep{\fill}}rlrr@{}}
\toprule
Rank & Systematic & $\chi^2_{\rm B}$ & $\delta\chi^2_{\rm B}$ \\
\midrule
1 & \texttt{sysHZComb1053} & 1327 & 36.78 \\
2 & \texttt{proc\_tb21}    & 1319 &  7.36 \\
3 & \texttt{sysHZComb1064} & 1315 &  3.80 \\
4 & \texttt{sysHZComb1114} & 1312 &  2.91 \\
5 & \texttt{sysHZComb1144} & 1310 &  2.53 \\
6 & \texttt{sysHZComb1111} & 1308 &  2.19 \\
7 & \texttt{sysHZComb1113} & 1306 &  1.88 \\
\midrule
8 (stop) & \texttt{proc\_nrl} & 1305 & 0.81 \\
\bottomrule
\end{tabular*}
\end{inlinetable}

In the final seven-source fit, the raw $\chi^2$ minimum falls from $1363$ to
$1313$, and the Bartlett correction gives $\chi^2_{\rm B}=1306$. With
$1131$ degrees of freedom, the corrected reduced $\chi^2$ is $1306/1131$, compared
with $1363/1131$ for the Gaussian baseline. The compatibility is substantially
improved, although the reduced $\chi^2$ remains above unity (the expectation for an
ideal description). 

To identify the kinematic regions that contribute most to the improvement
in fit quality, we examine the change in the data contribution
$\chi^2_{\rm data}$ between the baseline and errors-on-errors fits.
Here $\chi^2_{\rm data}$ denotes the data term defined by the first sum
in Eq.~(\ref{eq:xfitter_probability}), evaluated at the respective best fits
before the Bartlett correction, with the logarithmic normalisation and
NP constraint terms excluded. The largest contribution to its
$58$-unit decrease comes from the NC $e^+p$ sample at
$920~\mathrm{GeV}$ (377 fitted points), whose contribution falls
from $443$ to $418$. The reductions for the $820$,
$575$ and $460~\mathrm{GeV}$ samples are $7$, $8$ and $12$,
respectively; the NC $e^-p$ and CC samples also improve, though more moderately.

The distribution of this improvement in $Q^2$ and $x$ is shown in
Table~\ref{tab:region}, which gives
$\Delta\chi^2_{\rm data}=\chi^2_{\rm data}(\text{selected})-
\chi^2_{\rm data}(\text{baseline})$. Here, $\text{selected}$ denotes the final seven-source configuration.
Each populated cell gives the change followed, in parentheses, by the number of
measurements in the bin; negative values indicate a better description. The
largest contribution comes from the low-$Q^2$, low-$x$ region. The bins with
$Q^2<10~\mathrm{GeV}^2$ account for $-25$, of which $-21$ arises at
$x<10^{-3}$. As a complementary check of the low-$Q^2$ sensitivity seen in
Table~\ref{tab:region}, we varied the minimum $Q^2$ in the \mbox{HERAPDF2.0}, NNLO, RTOPT
configuration. Raising the cut from $3.5$ to $10~\mathrm{GeV}^2$ reduces the number of
selected systematic sources, as it excludes many data points sensitive
to errors-on-errors. Conversely, lowering it from $3.5$ to
$2~\mathrm{GeV}^2$ leaves the relative reduction in the overall $\chi^2$
largely unchanged and does not alter the conclusions of the
robustness study.

\begin{table*}[t]
\centering
\begin{minipage}{\objectwidthfraction\textwidth}
\centering
\caption{Change $\Delta\chi^2_{\rm data}$ in the pointwise data contribution
between the final seven-source errors-on-errors configuration and the
\mbox{HERAPDF2.0}, NNLO, RTOPT baseline fit, binned in $Q^2$ and $x$. Each populated cell gives this
change, followed in parentheses by the number of fitted measurements in the
bin.}
\label{tab:region}
\tablebodyfont
\renewcommand{\arraystretch}{1.25}
\setlength{\tabcolsep}{4pt}
\begin{tabular*}{\linewidth}{@{\extracolsep{\fill}}lrrrrr@{}}
\toprule
$Q^2\,[\mathrm{GeV}^2]$ & $x<10^{-3}$ & $10^{-3}\!\leq x<10^{-2}$ &
$10^{-2}\!\leq x<10^{-1}$ & $x\geq10^{-1}$ & total \\
\midrule
$<10$              & $-21.0$ (108) & $-2.5$ (17)  & $-1.6$ (4)   & --           & $-25.1$ \\
$10\leq Q^2<30$    & $-0.8$ (109)  & $-6.9$ (56)  & $+1.4$ (9)   & --           & $-6.2$ \\
$30\leq Q^2<100$   & $-7.0$ (27)   & $-5.2$ (128) & $-2.2$ (25)  & $+0.0$ (2)   & $-14.3$ \\
$100\leq Q^2<1000$ & --            & $+1.5$ (128) & $-2.6$ (195) & $-5.3$ (118) & $-6.4$ \\
$\geq1000$         & --            & --           & $-5.7$ (79)  & $+0.0$ (140) & $-5.7$ \\
\midrule
total               & $-28.7$       & $-13.1$      & $-10.7$      & $-5.2$       & $-57.7$ \\
\bottomrule
\end{tabular*}
\end{minipage}
\end{table*}

Finally, we assess how the reduction in $\chi^2_{\rm B}$ depends on the
assumed size of errors-on-errors. We keep the seven sources selected in
Table~\ref{tab:forward} fixed and vary their common $\epsilon$ from $0$ to $0.7$, refitting the PDF parameters and NPs
at each point, while all remaining sources retain $\epsilon_s=0$.
Figure~\ref{fig:hera_epsilon_scan} shows the corrected minimum
$\chi^2_{\rm B}$ as a function of $\epsilon$, recovering the Gaussian baseline at $\epsilon=0$. The reduction is appreciable already at $\epsilon\gtrsim0.1$ and slows beyond $\epsilon\approx0.3$.
However, $\chi^2_{\rm B}/n_{\rm dof}$ remains above unity throughout the range, indicating that errors-on-errors alone cannot account for the observed HERA tension.
Furthermore, repeating the forward selection at different $\epsilon$ values
would have little impact on the conclusions. The leading sources
remain unchanged, while the number selected increases with $\epsilon$ and the ordering of lower-ranked sources may vary slightly.

\begin{inlinefigure}
\centering
\includegraphics[width=\linewidth]{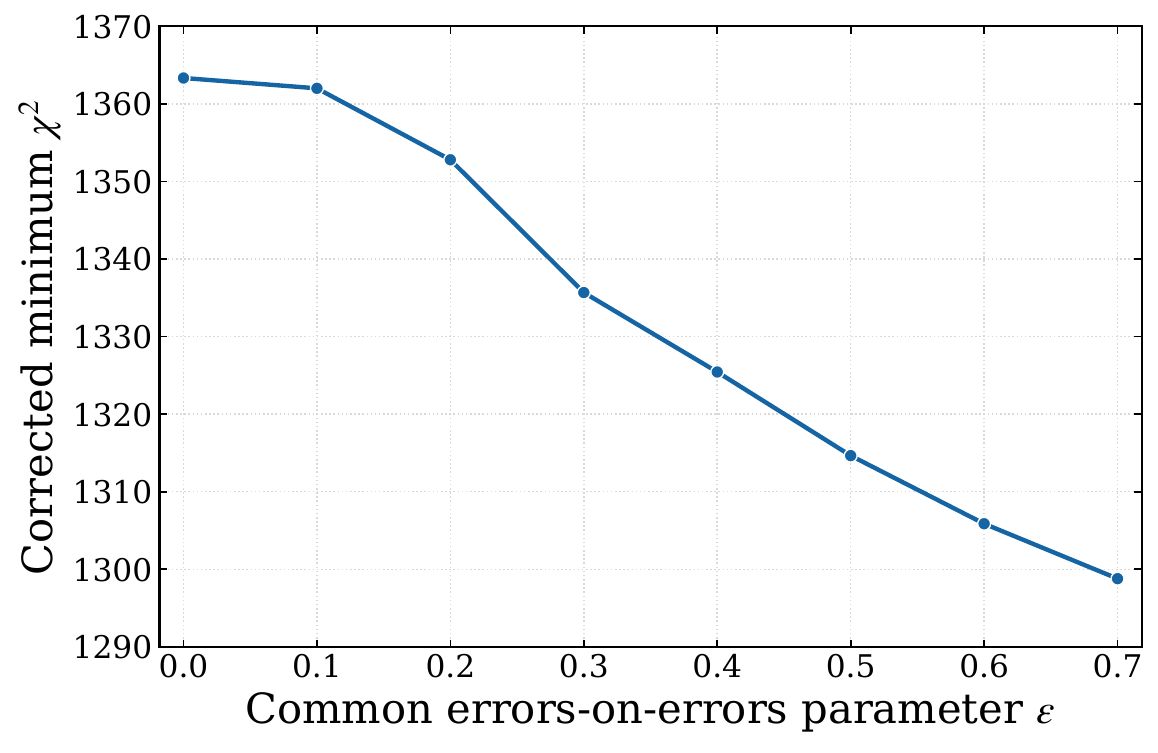}
\captionof{figure}{
Bartlett-corrected minimum $\chi^2_{\rm B}$ versus the common
errors-on-errors parameter $\epsilon$ for the seven sources
selected in Table~\ref{tab:forward}, using the
\mbox{HERAPDF2.0}, NNLO, RTOPT setup.}
\label{fig:hera_epsilon_scan}
\end{inlinefigure}

\subsection{Impact on PDFs and cross sections}
\label{sec:hera_pdf_impact}

\begin{figure*}[t]
\centering
\begin{minipage}{\objectwidthfraction\textwidth}
\centering
\includegraphics[width=\linewidth]{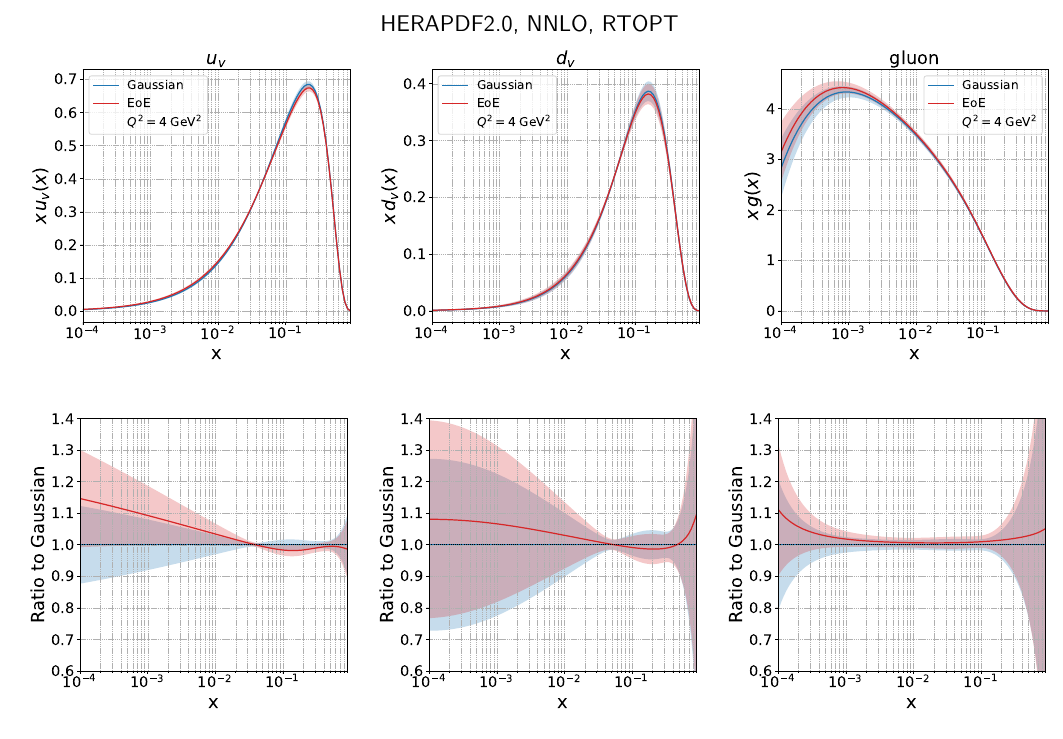}
\caption{Absolute PDFs and ratios for the \mbox{HERAPDF2.0}, NNLO, RTOPT Gaussian baseline and
the forward-selection errors-on-errors (EoE) fit at
$Q^2=4~\mathrm{GeV}^2$. From left to
right, the three columns show $xu_v$, $xd_v$, and $xg$, with the absolute
distributions in the upper row and their ratios to the baseline central PDF
in the lower row. }
\label{fig:hera_pdf_ratios}
\end{minipage}
\end{figure*}

We now turn to the impact of these systematic adjustments on the fitted
PDFs and cross-section predictions. Figure~\ref{fig:hera_pdf_ratios} compares
the final seven-source errors-on-errors fit with the \mbox{HERAPDF2.0}, NNLO, RTOPT baseline for $xu_v$, $xd_v$ and $xg$
at $Q^2=4~\mathrm{GeV}^2$, after evolution from the input scale. The central
values shift mainly at low $x$ but generally remain within the baseline
uncertainty bands; only the $u$-valence PDF lies marginally outside in some
regions. The confidence intervals broaden by up to about $25\%$ for $u_v$ and
$15\%$ for $d_v$ at low $x$, and by about $10$--$13\%$ for the gluon over most
of the high-$x$ region. Despite the  substantial reduction in $\chi^2_{\rm B}/n_{\rm dof}$, the PDFs remain broadly compatible with the baseline. The improvement in fit quality is accompanied by a broadening of the PDF confidence intervals, while the central values remain stable and show no large displacement.

\begin{figure*}[t]
\centering
\begin{minipage}{\objectwidthfraction\textwidth}
\centering
\includegraphics[width=\linewidth]{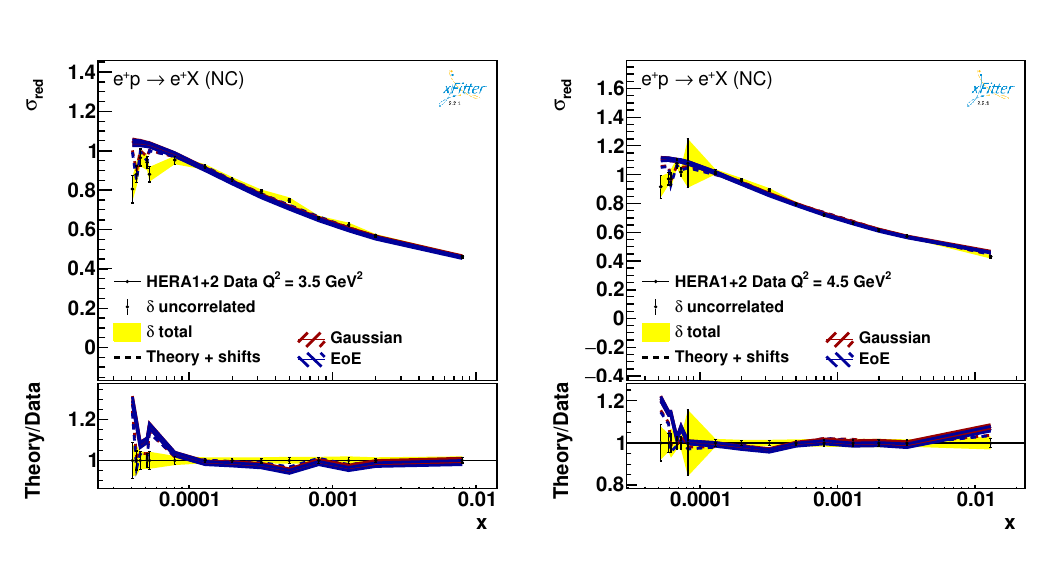}
\caption{Representative low-$Q^2$ reduced NC $e^+p$ cross sections at a
proton-beam energy of $920~\mathrm{GeV}$, at $Q^2=3.5~\mathrm{GeV}^2$ (left)
and $Q^2=4.5~\mathrm{GeV}^2$ (right). The \mbox{HERAPDF2.0}, NNLO, RTOPT Gaussian baseline and
forward-selection EoE fit are compared. The lower panels show the
predictions divided by the data; the dashed curves include the fitted
NP shifts, and the hatched bands are the $68.3\%$ Pumplin confidence
intervals.}
\label{fig:hera_xsec}
\end{minipage}
\end{figure*}

The corresponding cross sections show the same stability. Across all 1145
fitted measurements, the median point-by-point absolute change in the
unshifted theory prediction, i.e.\ the cross section obtained from the fitted
PDFs without accounting for NP shifts, is only $0.20\%$. This stability is
also present in the region giving the largest reduction in the data term
in Table~\ref{tab:region}, $Q^2<10~\mathrm{GeV}^2$ and $x<10^{-3}$.
Among its 108 measurements, the median change is $0.19\%$. These observations
suggest that errors-on-errors improve the fit through the cumulative effect
of small adjustments across many measurements, rather than large shifts
in a few cross sections. Figure~\ref{fig:hera_xsec} illustrates this stability
in the two lowest-$Q^2$ bins of the NC $e^+p$ sample, at $Q^2=3.5$ and
$4.5~\mathrm{GeV}^2$. At both scales, the two fits remain close in their
central predictions and in the widths of their confidence intervals.
The confidence-interval broadening is more pronounced in the CC $e^-p$
sample, which is more directly sensitive to the $u$-type valence
distributions: across that sample, the median increase in width is
about $9\%$.

Taken together, the errors-on-errors robustness test shows that the tension in the \mbox{HERAPDF2.0}, NNLO, RTOPT baseline is substantially reduced once the systematic-uncertainty model is given additional flexibility. The GVM allows the fit to accommodate small coherent fluctuations along the response patterns of the correlated sources, as expected when some systematic uncertainties are underestimated. This markedly improves data–theory compatibility without requiring comparably large shifts in the fitted PDFs or cross-section predictions. The stability of the central values reflects the nature of the tension in
this fit: the discrepancies are spread over many measurements, with no clear
outlier pulling the PDFs in a particular direction. The situation may be
different in global analyses combining different experiments and processes,
for instance when LHC data are added, where tensions between datasets or an
outlying dataset in a given kinematic region could lead errors-on-errors to
shift the central values as well, in the direction preferred by the mutually
consistent and more precise data, down-weighting outlying or less precise
datasets.

The conclusion of this section is limited to the 169 correlated sources, which account for about one quarter of the total uncertainty budget. The uncorrelated systematic components were aggregated before averaging and are not available source by source. The published data therefore do not allow us to assess the additional
effect of applying errors-on-errors to all systematic uncertainties
before the H1 and ZEUS measurements are combined.

\section{Interplay with the small-\texorpdfstring{$x$}{x} description}
\label{sec:interplay}

The largest errors-on-errors effects occur in the same low-$x$, low-$Q^2$
region in which the fit is particularly sensitive to the assumed PDF shape
and perturbative description.  We therefore consider separately the
additional flexibility of the input parametrisation and the perturbative
small-$x$ description, tested with matched fixed-order and NNLO+NLL$_x$
\mbox{FONLL-C} fits.  This allows us to test whether these changes replace, overlap
with, or remain complementary to the errors-on-errors treatment.

We first examine whether the errors-on-errors effect depends on the assumed PDF
functional form by considering the parametrisation proposed by Bonvini and
Giuli~\cite{Bonvini:2019wxf}, which we denote \mbox{HERAPDF-BG}. At the same starting scale
$Q_0^2=1.9~\mathrm{GeV}^2$, its generic form is
\begin{equation}
 \begin{aligned}
 xf(x,Q_0^2)={}&A_f x^{B_f}(1-x)^{C_f}\\
 &\times\Bigl[1+D_fx+E_fx^2\\
 &\qquad+F_f\ln x+G_f(\ln x)^2\Bigr].
 \end{aligned}
 \label{eq:bg_pdf_parametrisation}
\end{equation}
We choose these coefficients following Ref.~\cite{Bonvini:2019wxf}.
For the gluon, the terms proportional to $\ln x$ and $(\ln x)^2$,
with coefficients $F_g$ and $G_g$, replace the separate negative term
used in \mbox{HERAPDF2.0}. Both logarithmic terms are also included for $u_v$,
with coefficients $F_{u_v}$ and $G_{u_v}$. For the sea distributions,
we retain only the linear logarithmic terms, with a shared coefficient
$F_{\bar U}=F_{\bar D}$, and include an additional term
$D_{\bar D}x$ for $\bar D$. All remaining coefficient choices and constraints follow the prescription
of Sec.~\ref{sec:herafitsetup}.
The advantage of this parametrisation lies in the additional flexibility
provided by the logarithmic terms, which allow a wider range of PDF shapes
at small $x$ while vanishing as $x\to1$.
The resulting parametrisation has 18
fitted parameters, four more than the \mbox{HERAPDF2.0} form.

A distinct possibility is that part of the low-$x$ behaviour reflects limitations of the fixed-order perturbative description itself. At small $x$, contributions enhanced by powers of $\log(1/x)$ can become increasingly important and reduce the reliability of the fixed-order expansion~\cite{Ball:2017otu}.
The underlying leading-logarithmic small-$x$ (LL$_x$) results were developed in
Refs.~\cite{Kuraev:1976ge,Kuraev:1977fs,Balitsky:1978ic,Jaroszewicz:1982gr},
and the NLL$_x$ results in
Refs.~\cite{Catani:1994sq,Catani:1993rn,Fadin:1998py,Ciafaloni:1998gs}.
We follow Ref.~\cite{Bonvini:2019wxf} and resum these contributions using
the default NLL variant of the High-Energy Large Logarithms (HELL) code,
version~3.0~\cite{Bonvini:2016wki,Bonvini:2017ogt,Bonvini:2018xvt,Bonvini:2018ixe},
including their effects in the Dokshitzer--Gribov--Lipatov--Altarelli--Parisi
(DGLAP) splitting functions, the heavy-quark matching conditions and the
DIS coefficient functions. The calculation is matched to NNLO and
interfaced to \textsc{xFitter} through A PDF Evolution Library
(APFEL)~\cite{Bertone:2013vaa,Abdolmaleki:2018jln}. Since this implementation
uses the Fixed Order plus Next-to-Leading Logarithms (FONLL)
general-mass variable-flavour-number scheme~\cite{Forte:2010ta}, we adopt
its \mbox{FONLL-C} variant, appropriate at NNLO, and use an otherwise identical
fixed-order \mbox{FONLL-C} fit to isolate the effect of resummation.

With the \mbox{HERAPDF-BG} parametrisation and the original RTOPT setup of
Sec.~\ref{sec:herafitsetup}, we obtain
$\chi^2_{\rm min}/n_{\rm dof}=1301/1127$ under standard Gaussian constraints,
about $63$ units below the \mbox{HERAPDF2.0}, NNLO, RTOPT baseline. This reproduces the result
of Ref.~\cite{Bonvini:2019wxf}, where the clearest improvement occurs at
low $x$ and low $Q^2$. We refer to this setup as
\emph{\mbox{HERAPDF-BG}, NNLO, RTOPT}.
For the matched \mbox{FONLL-C} comparison, both fits retain the same parametrisation,
data and cuts, and $\alpha_s(M_Z)=0.118$, but use $Q_0=1.6~\mathrm{GeV}$,
$m_c=1.46~\mathrm{GeV}$ and $\mu_c=1.12m_c$, as in that reference.
We refer to this fixed-order reference as \emph{\mbox{HERAPDF-BG}, NNLO, \mbox{FONLL-C}}.
With these settings, it gives
$\chi^2_{\rm min}=1312$. Including NLL$_x$ resummation lowers this value
to $1283$.
We refer to this resummed setup as \emph{\mbox{HERAPDF-BG}, NNLO+NLL$_x$, \mbox{FONLL-C}}.
The \mbox{HERAPDF-BG}, NNLO, RTOPT and \mbox{HERAPDF-BG}, NNLO+NLL$_x$, \mbox{FONLL-C} Gaussian baselines
are used for the errors-on-errors study in the next subsection.

\subsection{Errors-on-errors analysis}
\label{sec:smallx_forward_selection}

We repeat the forward selection of Sec.~\ref{sec:robustness} for the
\mbox{HERAPDF-BG}, NNLO, RTOPT and \mbox{HERAPDF-BG}, NNLO+NLL$_x$, \mbox{FONLL-C} baselines, using the same $\epsilon_s=0.6$
assignment and retaining sources with $\delta\chi^2_{\rm B}>1$.
Tables~\ref{tab:forward_smallx} and~\ref{tab:forward_smallx_nllx} give the resulting sequences. Eight sources
are retained with the \mbox{HERAPDF-BG}, NNLO, RTOPT baseline.  All seven sources selected in the \mbox{HERAPDF2.0}, NNLO, RTOPT fit
are selected again, although in a different order, and the selection
additionally retains \texttt{sysHZComb1062}. In the \mbox{HERAPDF-BG}, NNLO+NLL$_x$, \mbox{FONLL-C} fit, the five retained sources are precisely the first five selected
with the \mbox{HERAPDF-BG}, NNLO, RTOPT baseline, although their marginal effects are changed by
the different theoretical description. This overlap nevertheless suggests
that the fits remain sensitive to many of the same systematic sources
when the low-$x$ parametrisation or theoretical description is changed. At the stopping point, errors-on-errors reduce
$\chi^2_{\rm B}/n_{\rm dof}$ from $1301/1127$ to $1254/1127$
for the \mbox{HERAPDF-BG}, NNLO, RTOPT fit, and from $1283/1127$
to $1247/1127$ for the \mbox{HERAPDF-BG}, NNLO+NLL$_x$, \mbox{FONLL-C} fit.

\begin{inlinetable}
\centering
\captionof{table}{Forward selection from the \mbox{HERAPDF-BG}, NNLO, RTOPT Gaussian baseline, with $\epsilon_s=0.6$. Entries give Bartlett-corrected minima and marginal reductions.}
\label{tab:forward_smallx}
\tablebodyfont
\renewcommand{\arraystretch}{1.25}
\setlength{\tabcolsep}{2.5pt}
\begin{tabular*}{\linewidth}{@{\extracolsep{\fill}}rlrr@{}}
\toprule
Rank & Systematic & $\chi^2_{\rm B}$ & $\delta\chi^2_{\rm B}$ \\
\midrule
1 & \texttt{sysHZComb1053} & 1270 & 30.69 \\
2 & \texttt{sysHZComb1062} & 1266 &  3.67 \\
3 & \texttt{sysHZComb1111} & 1263 &  3.02 \\
4 & \texttt{sysHZComb1114} & 1261 &  2.61 \\
5 & \texttt{sysHZComb1144} & 1258 &  2.62 \\
6 & \texttt{proc\_tb21}    & 1257 &  1.18 \\
7 & \texttt{sysHZComb1064} & 1255 &  1.69 \\
8 & \texttt{sysHZComb1113} & 1254 &  1.41 \\
\midrule
9 (stop) & \texttt{proc\_nrl} & 1253 & 0.78 \\
\bottomrule
\end{tabular*}
\end{inlinetable}

\begin{inlinetable}
\centering
\captionof{table}{Forward selection from the \mbox{HERAPDF-BG}, NNLO+NLL$_x$, \mbox{FONLL-C} Gaussian baseline, with $\epsilon_s=0.6$. Entries give Bartlett-corrected minima and marginal reductions.}
\label{tab:forward_smallx_nllx}
\tablebodyfont
\renewcommand{\arraystretch}{1.25}
\setlength{\tabcolsep}{2.5pt}
\begin{tabular*}{\linewidth}{@{\extracolsep{\fill}}rlrr@{}}
\toprule
Rank & Systematic & $\chi^2_{\rm B}$ & $\delta\chi^2_{\rm B}$ \\
\midrule
1 & \texttt{sysHZComb1053} & 1262 & 21.74 \\
2 & \texttt{sysHZComb1062} & 1255 &  7.02 \\
3 & \texttt{sysHZComb1111} & 1252 &  2.85 \\
4 & \texttt{sysHZComb1114} & 1250 &  2.29 \\
5 & \texttt{sysHZComb1144} & 1247 &  2.68 \\
\midrule
6 (stop) & \texttt{proc\_tb24} & 1246 & 0.99 \\
\bottomrule
\end{tabular*}
\end{inlinetable}

\begin{figure*}[t]
\centering
\begin{minipage}{\objectwidthfraction\textwidth}
\centering
\includegraphics[width=\linewidth]{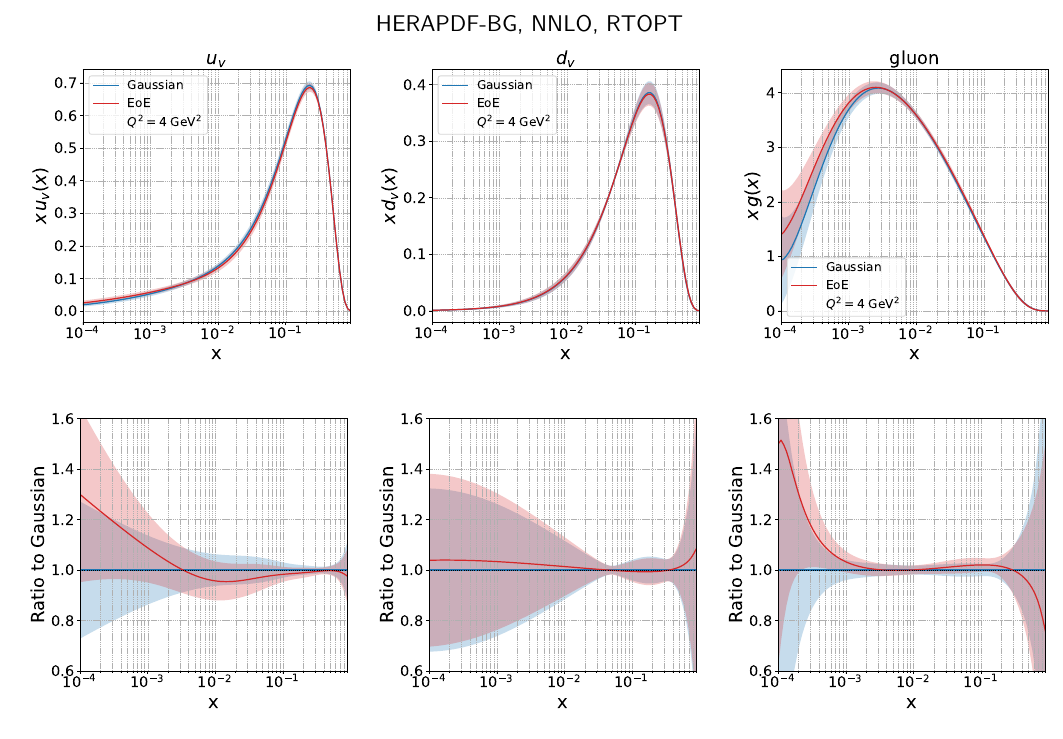}
\caption{PDFs and ratios for \mbox{HERAPDF-BG}, NNLO, RTOPT at $Q^2=4~\mathrm{GeV}^2$,
comparing the Gaussian baseline and forward-selection EoE fit. Columns show
$xu_v$, $xd_v$ and $xg$; ratios use the baseline central PDF.}
\label{fig:bg_pdf_ratios}
\end{minipage}
\end{figure*}

\begin{figure*}[t]
\centering
\begin{minipage}{\objectwidthfraction\textwidth}
\centering
\includegraphics[width=\linewidth]{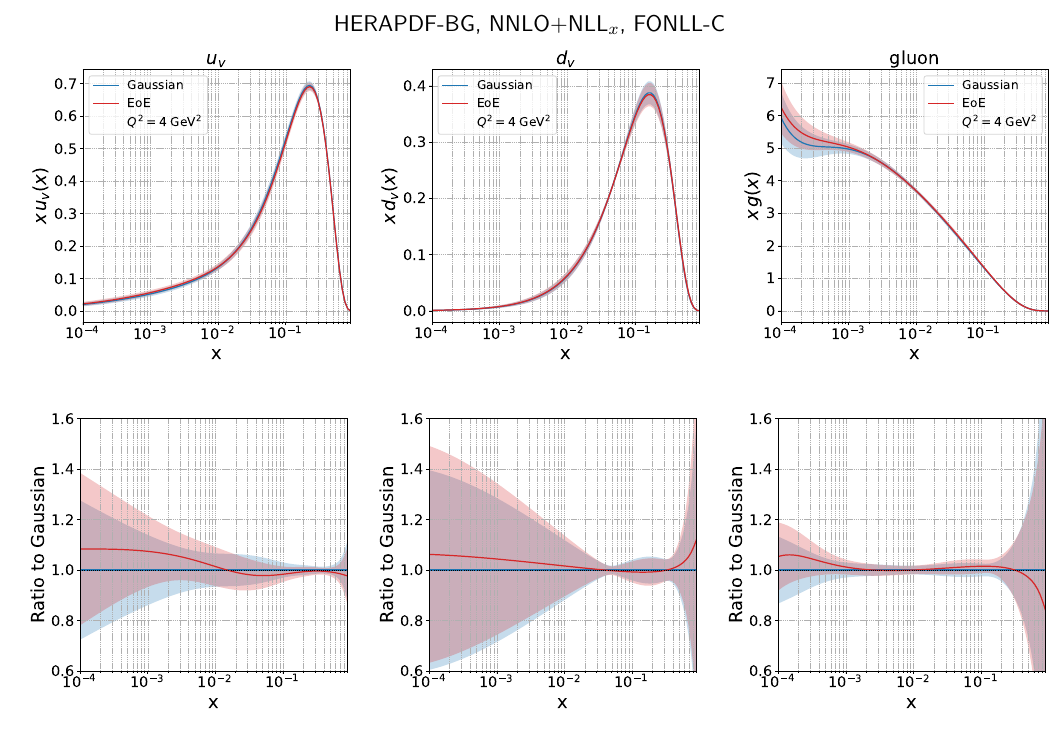}
\caption{Absolute PDFs and ratios for the \mbox{HERAPDF-BG}, NNLO+NLL$_x$, \mbox{FONLL-C} Gaussian baseline
and forward-selection EoE fit at $Q^2=4~\mathrm{GeV}^2$. From left to
right, the columns show $xu_v$, $xd_v$ and $xg$, with the absolute
distributions in the upper row and their ratios to the corresponding Gaussian
baseline central PDF in the lower row.}
\label{fig:smallx_eoe_pdf_ratios}
\end{minipage}
\end{figure*}

Figures~\ref{fig:bg_pdf_ratios} and~\ref{fig:smallx_eoe_pdf_ratios} compare
the Gaussian baselines with the corresponding final errors-on-errors fits
for $xu_v$, $xd_v$ and $xg$ at $Q^2=4~\mathrm{GeV}^2$. In the first configuration, the
central-value shifts are concentrated at the smallest $x$ and are most
visible in $u_v$ and the gluon, while $d_v$ remains comparatively stable.
The characteristic shapes generated by the additional parametrisation
freedom are preserved. The valence confidence intervals broaden in parts
of the $x$ range, most visibly for $u_v$ at small $x$, while the bands of
the two fits overlap throughout the displayed range. In the second configuration,
the central-value shifts are modest and remain within the corresponding
Gaussian confidence bands throughout the displayed range. The valence
confidence intervals broaden moderately in parts of the $x$ range, while
the low-$x$ gluon uncertainty is nearly unchanged. In both configurations,
the improvement in fit quality is therefore accompanied by
local central-value adjustments and a selective expansion of the PDF
confidence intervals.

For both configurations, the effect of errors-on-errors on the cross sections
is similar to that found in the \mbox{HERAPDF2.0}, NNLO, RTOPT study. Relative to the
corresponding Gaussian fit, the central predictions remain broadly
stable, with moderate changes in their confidence intervals.

Table~\ref{tab:stress_summary} summarises the forward-selection results for
all three fit configurations. The additional parametrisation freedom and
the change in theoretical treatment mainly affect the fitted low-$x$ PDF
shapes, whereas errors-on-errors provide additional freedom predominantly
through correlated NP shifts. The ordering of the fit minima is preserved
after introducing errors-on-errors. The substantial errors-on-errors
improvements that remain in all three configurations indicate that their
effect is not absorbed by the additional PDF flexibility or the
change in theoretical description.

\begin{inlinetable}
\centering
\captionof{table}{Summary of the forward-selection robustness tests. Each final fit assigns
$\epsilon_s=0.6$ to the selected sources and zero to all remaining sources.}
\label{tab:stress_summary}
\tablebodyfont
\renewcommand{\arraystretch}{1.25}
\setlength{\tabcolsep}{2.5pt}
\begin{tabular*}{\linewidth}{@{\extracolsep{\fill}}lrrr@{}}
\toprule
Fit setup & \shortstack[r]{Selected\\sources} & \shortstack[r]{Gaussian\\$\chi^2_{\rm min}$} & \shortstack[r]{Final\\$\chi^2_{\rm B}$} \\
\midrule
\parbox[c]{0.40\linewidth}{\raggedright \mbox{HERAPDF2.0},\\NNLO, RTOPT} & 7 & 1363 & 1306 \\
\addlinespace[6pt]
\parbox[c]{0.40\linewidth}{\raggedright \mbox{HERAPDF-BG},\\NNLO, RTOPT} & 8 & 1301 & 1254 \\
\addlinespace[6pt]
\parbox[c]{0.40\linewidth}{\raggedright \mbox{HERAPDF-BG},\\NNLO+NLL$_x$,\\\mbox{FONLL-C}} & 5 & 1283 & 1247 \\
\bottomrule
\end{tabular*}
\end{inlinetable}

\section{Illustrative informed application}
\label{sec:informed}

The tests described above help assess the robustness of the 
PDF fits
and identify systematic sources that merit closer investigation.
For a PDF determination intended for physics analyses,
each $\epsilon_s$ should be assigned a priori from information on
how the corresponding systematic uncertainty was constructed. Such
assignments are necessarily approximate: $\epsilon_s$ is more naturally
regarded as a scale reflecting the reliability of the quoted uncertainty
than as a precisely known parameter. In practice, a coarse classification with representative values such as
$\epsilon_s=0$, $0.3$ and $0.6$ can distinguish between systematic
uncertainties whose quoted sizes are regarded as well determined,
moderately uncertain or poorly determined, respectively.
Although approximate, such an assignment can 
represent an improvement relative to the Gaussian
treatment  ($\epsilon_s=0$ for all sources), provided 
the categories
reasonably reflect the reliability of the quoted uncertainties. The
Gaussian treatment is itself an approximation, imposing the more
restrictive and often unrealistic assumption that all systematic uncertainty sizes are known exactly. 

The task of assigning errors-on-errors is also simplified when most of the effect is concentrated in a
few sources. Reasonable estimates for these dominant sources may then
capture the main effect, without requiring equally detailed information for every systematic uncertainty.
Where the estimates remain uncertain, their impact can be made explicit
by presenting the relevant results as functions of the assumed
$\epsilon_s$, as illustrated for the fit minimum in
Fig.~\ref{fig:hera_epsilon_scan}.

For the HERA application, we adopt the following coarse classification.
We assign $\epsilon_s=0.6$ to the seven procedural uncertainties introduced
in the combination, which quantify its dependence on the treatment of
systematic uncertainties as additive or multiplicative, assumptions on
correlations between data sets, and four variations derived from groups of
sources with large pulls; see Sec.~4.5 of Ref.~\cite{H1:2015ubc}.
Their estimation from discrete alternative procedures limits the precision
of their assigned sizes, motivating a value close to the
two-point-systematic benchmark $\epsilon_s=1/\sqrt{2}\simeq0.7$. Then, we assign $\epsilon_s=0.3$ to the 40 \texttt{sysHZComb} sources
identified by the HERA combination pull study, discussed in
Sec.~4.5.3 and Fig.~3 of Ref.~\cite{H1:2015ubc} and explicitly
listed in Sec.~3.4.3 (p.~57; see also Figs.~3.5--3.6) of
Ref.~\cite{Turkot:2017thesis}. Their normalised combination pulls,
defined in Eq.~(3.12) of that reference, exceed $1.3$ in absolute value.
As noted in Ref.~\cite{H1:2015ubc}, such large pulls might indicate that these
uncertainties were underestimated or do not fulfil the assumptions of the
linear averaging procedure. Their quoted sizes are therefore less reliable
than those of the remaining sources, motivating the intermediate value
$\epsilon_s=0.3$.
These pulls are evaluated using all inclusive NC and CC measurements
simultaneously over the full published $Q^2$ and $x$ coverage, before
imposing the $Q^2$ cut of our PDF fits. We therefore use the sources identified in the published HERA combination study, without adjusting the selection to our PDF-fit results.
The selected sources span very-low-$Q^2$ HERA~I data, reduced-energy HERA~II
data, medium- and high-$Q^2$ data, and normalisation effects.
The remaining 122 sources retain $\epsilon_s=0$.\footnote{Using the
terminology introduced by Pekka Sinervo, we call systematic errors with
$\epsilon_s=0.0$, $0.3$ and $0.6$ good, bad and ugly~\cite{Sinervo:2003wm}.}

\begin{figure*}[t]
\centering
\begin{minipage}{\objectwidthfraction\textwidth}
\centering
\includegraphics[width=\linewidth]{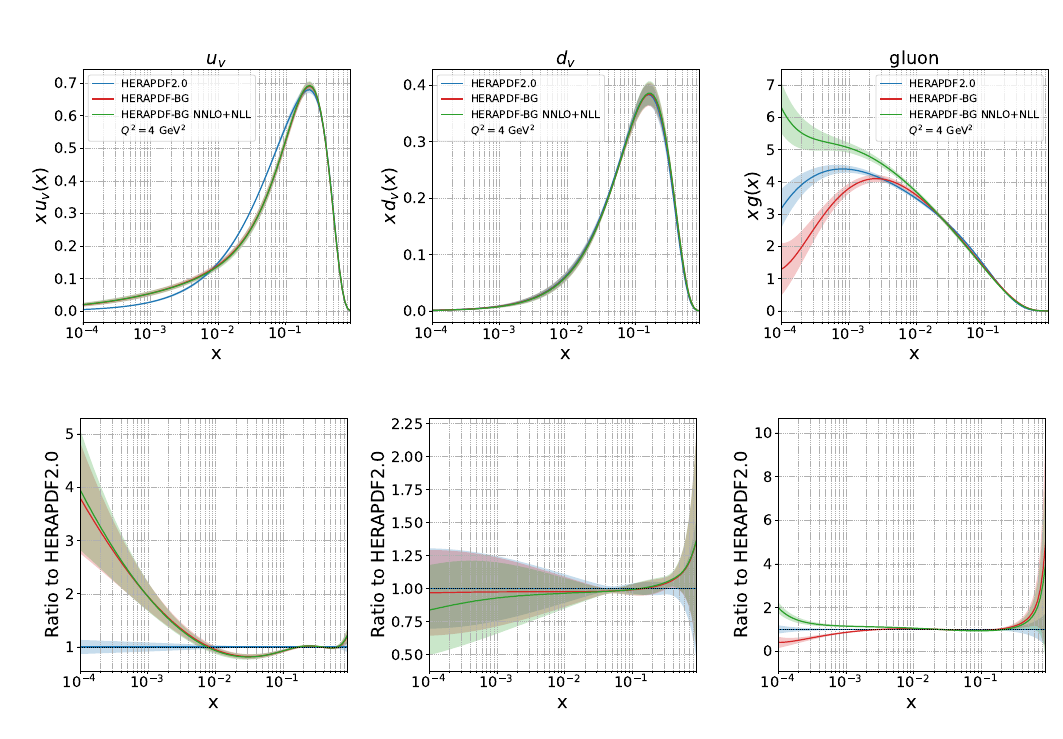}
\caption{Comparison of the informed errors-on-errors PDF sets at $Q^2=4~\mathrm{GeV}^2$
for \mbox{HERAPDF2.0}, NNLO, RTOPT; \mbox{HERAPDF-BG}, NNLO, RTOPT; and
\mbox{HERAPDF-BG}, NNLO+NLL$_x$, \mbox{FONLL-C}. All three fits use the same
$\epsilon_s$ assignment described in the text. From left to right,
the columns show $xu_v$, $xd_v$ and $xg$, with absolute distributions
and their confidence bands in the upper row and ratios to the informed
\mbox{HERAPDF2.0}, NNLO, RTOPT central PDFs in the lower row.}
\label{fig:informed_pdf_comparison}
\end{minipage}
\end{figure*}

We apply the $\epsilon_s$ assignment described above to the three fit
setups in Table~\ref{tab:informed_results},
refitting all PDF parameters and NPs in each case.
The informed configurations give $\chi^2_{\rm B}/n_{\rm dof}$ equal to
$1328/1131$, $1276/1127$ and $1267/1127$, compared with $1363/1131$,
$1301/1127$ and $1283/1127$ for the respective Gaussian baselines.
The informed treatment therefore has a significant impact on data--theory
compatibility in all three setups, as expected: several of the procedural and
pull-study sources, in particular \texttt{sysHZComb1053}, were also identified
by the forward selections of Secs.~\ref{sec:robustness}
and~\ref{sec:smallx_forward_selection}. Figure~\ref{fig:informed_pdf_comparison}
compares the PDF sets obtained with the informed errors-on-errors assignment
at $Q^2=4~\mathrm{GeV}^2$. At small $x$, the two fits with the \mbox{HERAPDF-BG} parametrisation yield
$u_v$ distributions whose confidence bands remain clearly separated from
the \mbox{HERAPDF2.0}, NNLO, RTOPT band.
In the same region, the three gluon distributions have non-overlapping
confidence bands.

\begin{inlinetable}
\centering
\captionof{table}{Goodness of fit for the three Gaussian baselines and the corresponding
informed errors-on-errors configurations.}
\label{tab:informed_results}
\tablebodyfont
\renewcommand{\arraystretch}{1.25}
\setlength{\tabcolsep}{2.5pt}
\begin{tabular*}{\linewidth}{@{\extracolsep{\fill}}lrrr@{}}
\toprule
Fit setup & $n_{\rm dof}$ & \shortstack[r]{Gaussian\\$\chi^2_{\rm min}$} & \shortstack[r]{Informed\\$\chi^2_{\rm B}$} \\
\midrule
\parbox[c]{0.40\linewidth}{\raggedright \mbox{HERAPDF2.0},\\NNLO, RTOPT} & 1131 & 1363 & 1328 \\
\addlinespace[6pt]
\parbox[c]{0.40\linewidth}{\raggedright \mbox{HERAPDF-BG},\\NNLO, RTOPT} & 1127 & 1301 & 1276 \\
\addlinespace[6pt]
\parbox[c]{0.40\linewidth}{\raggedright \mbox{HERAPDF-BG},\\NNLO+NLL$_x$,\\\mbox{FONLL-C}} & 1127 & 1283 & 1267 \\
\bottomrule
\end{tabular*}
\end{inlinetable}

This exercise illustrates how experimental information can guide
the assignment of errors-on-errors in a PDF fit.
Here, the chosen values of $\epsilon_s$ are indicative but entirely realistic; future analyses
of new datasets could assign $\epsilon_s$ more precisely using detailed
information on how each systematic uncertainty was estimated. The configuration needed to reproduce the informed
\mbox{HERAPDF2.0}, NNLO, RTOPT fit is made available
in the \texttt{errors-on-errors} folder of the \textsc{xFitter}
\href{https://gitlab.cern.ch/fitters/xfitter/-/tree/master/examples}
{examples directory}.

\section{Conclusions}
\label{sec:conclusions}

PDF fits often yield values of $\chi^2/n_{\rm dof}$ significantly
larger than unity, indicating residual tensions that may arise not only from
the theory or the PDF parametrisation, but also from the experimental
uncertainty model.  In this work, we have investigated, using the combined
HERA inclusive dataset, to what extent this discrepancy can be attributed to
uncertainty in the assigned sizes of systematic errors.  We have
done so by modelling uncertainties in the assignment of systematic errors,
informally referred to as errors-on-errors, with the Gamma Variance Model
implemented in \textsc{xFitter}.

To quantify the largest errors-on-errors effect accessible within the combined
HERA dataset, we first performed a robustness study of the fit.  The study
uses forward selection of correlated systematic sources at a common stress
value $\epsilon_s=0.6$ to probe the sensitivity of the fit to errors-on-errors.
By identifying the systematic sources whose errors-on-errors most strongly
affect the fit, the study can also help prioritise experimental follow-up to
reassess their uncertainty estimates.
The forward selection identifies seven systematic sources with appreciable
errors-on-errors effects in the \mbox{HERAPDF2.0}, NNLO, RTOPT baseline, which follows
Ref.~\cite{H1:2015ubc}, reducing $\chi^2_{\rm B}/n_{\rm dof}$ from
$1363/1131$ to $1306/1131$, with the improvement driven particularly by
the data at low $Q^2$ and low $x$.
Despite this substantial improvement, the PDF central values remain broadly
stable, while the confidence intervals expand moderately. The Gamma Variance Model thus improves the inferred data--theory
compatibility by allowing greater freedom in the coherent shifts associated
with correlated systematic sources, without requiring large changes in the
fitted PDFs or individual cross-section predictions. This behaviour is
consistent with some of the quoted systematic uncertainties being
underestimated.

Since both the more flexible \mbox{HERAPDF-BG} parametrisation and small-$x$
resummation improve the description of the same low-$x$, low-$Q^2$ region
where the errors-on-errors effect is largest, it is natural to ask how much
of this effect remains once these improvements are included. Repeating
the study with the \mbox{HERAPDF-BG}, NNLO, RTOPT and \mbox{HERAPDF-BG}, NNLO+NLL$_x$, \mbox{FONLL-C}
configurations still yields substantial reductions in
$\chi^2_{\rm B}$. The effect is therefore not absorbed by either the
additional PDF-shape flexibility or the improved perturbative description.

While the forward selection estimates the largest errors-on-errors effect accessible
in the present setup, a PDF determination including errors-on-errors should
assign the $\epsilon_s$ values a priori, using information on how the various sources of uncertainty were quantified.
As an illustration, we use the published HERA studies to classify the systematic
sources: the seven procedural uncertainties are assigned $\epsilon_s=0.6$,
the 40 correlated sources highlighted by the HERA pull study are assigned
$\epsilon_s=0.3$, and all remaining sources retain $\epsilon_s=0$.  The resulting fit minima and PDF distributions are presented in
Table~\ref{tab:informed_results} and Fig.~\ref{fig:informed_pdf_comparison}, respectively.  Although illustrative,
these results show that a transparent, externally motivated errors-on-errors
treatment improves the goodness of fit across all three fit setups while
preserving the characteristic low-$x$ differences in the $u_v$ and gluon
distributions.

Taken together, these studies show that, within a fixed PDF parametrisation
and theoretical setup, errors-on-errors improve the inferred data--theory
compatibility and increase the uncertainties on fitted parameters,
without substantially changing the central PDF and cross-section
predictions relative to the corresponding Gaussian fits. This stability
may not persist in global fits combining different experiments and processes,
where datasets may be in tension: errors-on-errors could then shift the central
values as well, by down-weighting outlying or less precise datasets in favour
of the mutually consistent and more precise ones.

\pdfbookmark[1]{Acknowledgments}{acknowledgments}
\bmhead*{Acknowledgments}
The authors are grateful to Robert Thorne, Matthew Reader,
Amanda Cooper-Sarkar, Davide Malito and Gitanjali Poddar for helpful
discussions.
E. C. and G. C. thank the UK Science and Technology Facilities Council
for its support.
The work of O. Z. has been supported by the \emph{MSCA4Ukraine Programme}
of the European Commission through the Alexander von Humboldt Foundation.
\par

\nocite{H1:2015ubc,Bonvini:2019wxf}

\bibliography{references}

@article{H1:2015ubc,
  author        = {{H1 and ZEUS Collaborations}},
  title         = {Combination of Measurements of Inclusive Deep Inelastic
                   $e^{\pm}p$ Scattering Cross Sections and {QCD} Analysis of
                   {HERA} Data},
  eprint        = {1506.06042},
  archiveprefix = {arXiv},
  primaryclass  = {hep-ex},
  journal       = {Eur. Phys. J. C},
  volume        = {75},
  number        = {12},
  pages         = {580},
  year          = {2015},
  doi           = {10.1140/epjc/s10052-015-3710-4}
}

@phdthesis{Turkot:2017thesis,
  author        = {Turkot, Oleksii},
  title         = {Search for contact interactions and finite quark radius
                   using combined {$e^{\pm}p$} data at {HERA}},
  school        = {Universit{\"a}t Hamburg},
  year          = {2017},
  note          = {DESY-THESIS-2017-035},
  doi           = {10.3204/PUBDB-2017-09190}
}

@article{ATLAS:2016WZPDF,
  author        = {{ATLAS Collaboration}},
  title         = {Precision Measurement and Interpretation of Inclusive
                   {$W^+$}, {$W^-$} and {$Z/\gamma^*$} Production Cross Sections
                   with the {ATLAS} Detector},
  eprint        = {1612.03016},
  archiveprefix = {arXiv},
  primaryclass  = {hep-ex},
  journal       = {Eur. Phys. J. C},
  volume        = {77},
  number        = {6},
  pages         = {367},
  year          = {2017},
  doi           = {10.1140/epjc/s10052-017-4911-9}
}

@article{ATLAS:2021VJetsPDF,
  author        = {{ATLAS Collaboration}},
  title         = {Determination of the Parton Distribution Functions of the
                   Proton from {ATLAS} Measurements of Differential {$W^\pm$}
                   and {$Z$} Boson Production in Association with Jets},
  eprint        = {2101.05095},
  archiveprefix = {arXiv},
  primaryclass  = {hep-ex},
  journal       = {JHEP},
  volume        = {07},
  pages         = {223},
  year          = {2021},
  doi           = {10.1007/JHEP07(2021)223}
}

@article{ATLAS:2022PDF21,
  author        = {{ATLAS Collaboration}},
  title         = {Determination of the Parton Distribution Functions of the
                   Proton Using Diverse {ATLAS} Data from $pp$ Collisions at
                   $\sqrt{s}=7$, 8 and 13 {TeV}},
  eprint        = {2112.11266},
  archiveprefix = {arXiv},
  primaryclass  = {hep-ex},
  journal       = {Eur. Phys. J. C},
  volume        = {82},
  number        = {5},
  pages         = {438},
  year          = {2022},
  doi           = {10.1140/epjc/s10052-022-10217-z}
}

@article{Bonvini:2019wxf,
  author        = {Bonvini, Marco and Giuli, Francesco},
  title         = {A new simple {PDF} parametrization: improved description of
                   the {HERA} data},
  eprint        = {1902.11125},
  archiveprefix = {arXiv},
  primaryclass  = {hep-ph},
  journal       = {Eur. Phys. J. Plus},
  volume        = {134},
  number        = {10},
  pages         = {531},
  year          = {2019},
  doi           = {10.1140/epjp/i2019-12872-x}
}

@article{Bertone:2013vaa,
  author        = {Bertone, Valerio and Carrazza, Stefano and Rojo, Juan},
  title         = {{APFEL}: A {PDF} Evolution Library with {QED} corrections},
  eprint        = {1310.1394},
  archiveprefix = {arXiv},
  primaryclass  = {hep-ph},
  journal       = {Comput. Phys. Commun.},
  volume        = {185},
  pages         = {1647--1668},
  year          = {2014},
  doi           = {10.1016/j.cpc.2014.03.007}
}

@article{Kuraev:1976ge,
  author        = {Kuraev, E. A. and Lipatov, L. N. and Fadin, Victor S.},
  title         = {Multi-{Reggeon} Processes in the {Yang--Mills} Theory},
  journal       = {Sov. Phys. JETP},
  volume        = {44},
  pages         = {443--450},
  year          = {1976}
}

@article{Kuraev:1977fs,
  author        = {Kuraev, E. A. and Lipatov, L. N. and Fadin, Victor S.},
  title         = {The {Pomeranchuk} Singularity in Nonabelian Gauge Theories},
  journal       = {Sov. Phys. JETP},
  volume        = {45},
  pages         = {199--204},
  year          = {1977}
}

@article{Balitsky:1978ic,
  author        = {Balitsky, I. I. and Lipatov, L. N.},
  title         = {The {Pomeranchuk} Singularity in Quantum Chromodynamics},
  journal       = {Sov. J. Nucl. Phys.},
  volume        = {28},
  pages         = {822--829},
  year          = {1978}
}

@article{Jaroszewicz:1982gr,
  author        = {Jaroszewicz, T.},
  title         = {Gluonic {Regge} Singularities and Anomalous Dimensions in
                   {QCD}},
  reportnumber  = {INP-1178/PH},
  journal       = {Phys. Lett. B},
  volume        = {116},
  pages         = {291--294},
  year          = {1982},
  doi           = {10.1016/0370-2693(82)90345-8}
}

@article{Catani:1994sq,
  author        = {Catani, S. and Hautmann, F.},
  title         = {High-energy factorization and small $x$ deep inelastic
                   scattering beyond leading order},
  eprint        = {hep-ph/9405388},
  archiveprefix = {arXiv},
  reportnumber  = {CAVENDISH-HEP-94-01},
  journal       = {Nucl. Phys. B},
  volume        = {427},
  pages         = {475--524},
  year          = {1994},
  doi           = {10.1016/0550-3213(94)90636-X}
}

@article{Catani:1993rn,
  author        = {Catani, S. and Hautmann, F.},
  title         = {Quark anomalous dimensions at small $x$},
  reportnumber  = {CERN-TH-6910-93},
  journal       = {Phys. Lett. B},
  volume        = {315},
  pages         = {157--163},
  year          = {1993},
  doi           = {10.1016/0370-2693(93)90174-G}
}

@article{Fadin:1998py,
  author        = {Fadin, Victor S. and Lipatov, L. N.},
  title         = {{BFKL} pomeron in the next-to-leading approximation},
  eprint        = {hep-ph/9802290},
  archiveprefix = {arXiv},
  reportnumber  = {DESY-98-033},
  journal       = {Phys. Lett. B},
  volume        = {429},
  pages         = {127--134},
  year          = {1998},
  doi           = {10.1016/S0370-2693(98)00473-0}
}

@article{Ciafaloni:1998gs,
  author        = {Ciafaloni, Marcello and Camici, Gianni},
  title         = {Energy scale(s) and next-to-leading {BFKL} equation},
  eprint        = {hep-ph/9803389},
  archiveprefix = {arXiv},
  reportnumber  = {DFF-302-03-98},
  journal       = {Phys. Lett. B},
  volume        = {430},
  pages         = {349--354},
  year          = {1998},
  doi           = {10.1016/S0370-2693(98)00551-6}
}

@article{Bonvini:2016wki,
  author        = {Bonvini, Marco and Marzani, Simone and Peraro, Tiziano},
  title         = {Small-$x$ resummation from {HELL}},
  eprint        = {1607.02153},
  archiveprefix = {arXiv},
  primaryclass  = {hep-ph},
  journal       = {Eur. Phys. J. C},
  volume        = {76},
  number        = {11},
  pages         = {597},
  doi           = {10.1140/epjc/s10052-016-4445-6},
  year          = {2016}
}

@article{Bonvini:2017ogt,
  author        = {Bonvini, Marco and Marzani, Simone and Muselli, Claudio},
  title         = {Towards parton distribution functions with small-$x$
                   resummation: {HELL} 2.0},
  eprint        = {1708.07510},
  archiveprefix = {arXiv},
  primaryclass  = {hep-ph},
  journal       = {JHEP},
  volume        = {12},
  pages         = {117},
  doi           = {10.1007/JHEP12(2017)117},
  year          = {2017}
}

@article{Bonvini:2018ixe,
  author        = {Bonvini, Marco},
  title         = {Small-$x$ phenomenology at the {LHC} and beyond:
                   {HELL} 3.0 and the case of the {Higgs} cross section},
  eprint        = {1805.08785},
  archiveprefix = {arXiv},
  primaryclass  = {hep-ph},
  journal       = {Eur. Phys. J. C},
  volume        = {78},
  number        = {10},
  pages         = {834},
  doi           = {10.1140/epjc/s10052-018-6315-x},
  year          = {2018}
}

@article{Ball:2017otu,
  author        = {Ball, Richard D. and Bertone, Valerio and Bonvini, Marco and
                   Marzani, Simone and Rojo, Juan and Rottoli, Luca},
  title         = {Parton distributions with small-$x$ resummation: evidence
                   for {BFKL} dynamics in {HERA} data},
  eprint        = {1710.05935},
  archiveprefix = {arXiv},
  primaryclass  = {hep-ph},
  journal       = {Eur. Phys. J. C},
  volume        = {78},
  number        = {4},
  pages         = {321},
  doi           = {10.1140/epjc/s10052-018-5774-4},
  year          = {2018}
}

@article{Forte:2010ta,
  author        = {Forte, Stefano and Laenen, Eric and Nason, Paolo and
                   Rojo, Juan},
  title         = {Heavy quarks in deep-inelastic scattering},
  eprint        = {1001.2312},
  archiveprefix = {arXiv},
  primaryclass  = {hep-ph},
  journal       = {Nucl. Phys. B},
  volume        = {834},
  pages         = {116--162},
  doi           = {10.1016/j.nuclphysb.2010.03.014},
  year          = {2010}
}

@article{Alekhin:2014irh,
  author        = {Alekhin, Sergey and others},
  title         = {{HERAFitter}, Open Source {QCD} Fit Project},
  eprint        = {1410.4412},
  archiveprefix = {arXiv},
  primaryclass  = {hep-ph},
  journal       = {Eur. Phys. J. C},
  volume        = {75},
  number        = {7},
  pages         = {304},
  year          = {2015},
  doi           = {10.1140/epjc/s10052-015-3480-z}
}

@article{Bertone:2017tig,
  author        = {Bertone, Valerio and others},
  collaboration = {{xFitter Developers' Team}},
  title         = {{xFitter 2.0.0}: An Open Source {QCD} Fit Framework},
  eprint        = {1709.01151},
  archiveprefix = {arXiv},
  primaryclass  = {hep-ph},
  journal       = {PoS},
  volume        = {DIS2017},
  pages         = {203},
  year          = {2018},
  doi           = {10.22323/1.297.0203}
}

@misc{xFitter:2022zjb,
  author        = {Abdolmaleki, H. and others},
  collaboration = {xFitter},
  title         = {{xFitter}: An Open Source {QCD} Analysis Framework.
                   {A} resource and reference document for the {Snowmass} study},
  eprint        = {2206.12465},
  archiveprefix = {arXiv},
  primaryclass  = {hep-ph},
  year          = {2022},
  note          = {arXiv:2206.12465 [hep-ph]}
}

@article{Botje:2010ay,
  author        = {Botje, Michiel},
  title         = {{QCDNUM}: Fast {QCD} Evolution and Convolution},
  eprint        = {1005.1481},
  archiveprefix = {arXiv},
  primaryclass  = {hep-ph},
  journal       = {Comput. Phys. Commun.},
  volume        = {182},
  pages         = {490--532},
  year          = {2011},
  doi           = {10.1016/j.cpc.2010.10.020}
}

@article{Thorne:1997ga,
  author        = {Thorne, Robert S. and Roberts, R. G.},
  title         = {An Ordered Analysis of Heavy Flavor Production in Deep
                   Inelastic Scattering},
  eprint        = {hep-ph/9709442},
  archiveprefix = {arXiv},
  primaryclass  = {hep-ph},
  journal       = {Phys. Rev. D},
  volume        = {57},
  pages         = {6871--6898},
  year          = {1998},
  doi           = {10.1103/PhysRevD.57.6871}
}

@article{Thorne:2006qt,
  author        = {Thorne, Robert S.},
  title         = {A Variable-Flavor Number Scheme for {NNLO}},
  eprint        = {hep-ph/0601245},
  archiveprefix = {arXiv},
  primaryclass  = {hep-ph},
  journal       = {Phys. Rev. D},
  volume        = {73},
  pages         = {054019},
  year          = {2006},
  doi           = {10.1103/PhysRevD.73.054019}
}

@article{Thorne:2012az,
  author        = {Thorne, Robert S.},
  title         = {Effect of Changes of Variable Flavour Number Scheme on
                   Parton Distribution Functions and Predicted Cross
                   Sections},
  eprint        = {1201.6180},
  archiveprefix = {arXiv},
  primaryclass  = {hep-ph},
  journal       = {Phys. Rev. D},
  volume        = {86},
  pages         = {074017},
  year          = {2012},
  doi           = {10.1103/PhysRevD.86.074017}
}

@article{Pumplin:2001ct,
  author        = {Pumplin, J. and Stump, D. and Brock, R. and Casey, D. and
                   Huston, J. and Kalk, J. and Lai, H. L. and Tung, W. K.},
  title         = {Uncertainties of Predictions from Parton Distribution
                   Functions. {II}. {The} {Hessian} Method},
  eprint        = {hep-ph/0101032},
  archiveprefix = {arXiv},
  journal       = {Phys. Rev. D},
  volume        = {65},
  pages         = {014013},
  year          = {2001},
  doi           = {10.1103/PhysRevD.65.014013}
}

@article{Stump:2001gu,
  author        = {Stump, D. and Pumplin, J. and Brock, R. and Casey, D. and
                   Huston, J. and Kalk, J. and Lai, H. L. and Tung, W. K.},
  title         = {Uncertainties of Predictions from Parton Distribution
                   Functions. {I}. {The} {Lagrange} Multiplier Method},
  eprint        = {hep-ph/0101051},
  archiveprefix = {arXiv},
  journal       = {Phys. Rev. D},
  volume        = {65},
  pages         = {014012},
  year          = {2001},
  doi           = {10.1103/PhysRevD.65.014012}
}

@article{Cowan:2018gqq,
  author        = {Cowan, Glen},
  title         = {Statistical Models with Uncertain Error Parameters},
  eprint        = {1809.05778},
  archiveprefix = {arXiv},
  primaryclass  = {physics.data-an},
  journal       = {Eur. Phys. J. C},
  volume        = {79},
  number        = {2},
  pages         = {133},
  year          = {2019},
  doi           = {10.1140/epjc/s10052-019-6644-4}
}

@article{Canonero:2024zlb,
  author        = {Canonero, Enzo and Cowan, Glen},
  title         = {Correlated systematic uncertainties and errors-on-errors in
                   measurement combinations with an application to the
                   7--8 {TeV} {ATLAS--CMS} top quark mass combination},
  eprint        = {2407.05322},
  archiveprefix = {arXiv},
  primaryclass  = {physics.data-an},
  journal       = {Eur. Phys. J. C},
  volume        = {85},
  number        = {2},
  pages         = {156},
  year          = {2025},
  doi           = {10.1140/epjc/s10052-025-13884-w}
}

@article{Canonero:2023qiz,
  author        = {Canonero, Enzo and Brazzale, Alessandra Rosalba and
                   Cowan, Glen},
  title         = {Higher-order asymptotic corrections and their application
                   to the {Gamma Variance Model}},
  eprint        = {2304.10574},
  archiveprefix = {arXiv},
  primaryclass  = {physics.data-an},
  journal       = {Eur. Phys. J. C},
  volume        = {83},
  pages         = {1100},
  year          = {2023},
  doi           = {10.1140/epjc/s10052-023-12263-7}
}

@article{Bartlett:1937,
  author  = {Bartlett, M. S.},
  title   = {Properties of Sufficiency and Statistical Tests},
  journal = {Proc. Roy. Soc. Lond. A},
  volume  = {160},
  pages   = {268--282},
  year    = {1937},
  doi     = {10.1098/rspa.1937.0109}
}

@article{Lawley:1956,
  author  = {Lawley, D. N.},
  title   = {A General Method for Approximating to the Distribution of
             Likelihood Ratio Criteria},
  journal = {Biometrika},
  volume  = {43},
  number  = {3/4},
  pages   = {295--303},
  year    = {1956},
  doi     = {10.2307/2332908}
}

@article{Cordeiro:1993,
  author  = {Cordeiro, Gauss M.},
  title   = {General Matrix Formulae for Computing {Bartlett} Corrections},
  journal = {Stat. Probab. Lett.},
  volume  = {16},
  number  = {1},
  pages   = {11--18},
  year    = {1993},
  doi     = {10.1016/0167-7152(93)90115-Y}
}

@book{CordeiroCribariNeto:2014,
  author    = {Cordeiro, Gauss M. and Cribari-Neto, Francisco},
  title     = {An Introduction to {Bartlett} Correction and Bias Reduction},
  publisher = {Springer},
  address   = {Berlin, Heidelberg},
  year      = {2014},
  doi       = {10.1007/978-3-642-55255-7}
}

@article{Sinervo:2003wm,
    author = "Sinervo, Pekka",
    editor = "Lyons, L. and Mount, R. P. and Reitmeyer, R.",
    title = "{Definition and Treatment of Systematic Uncertainties in High Energy Physics and Astrophysics}",
    reportNumber = "PHYSTAT-2003-TUAT004",
    journal = "eConf",
    volume = "C030908",
    pages = "122--129",
    year = "2003"
}

@article{Abdolmaleki:2018jln,
    author         = "Abdolmaleki, Hamed and others",
    collaboration  = "xFitter Developers' Team",
    title          = "{Impact of low-$x$ resummation on {QCD} analysis of {HERA} data}",
    eprint         = "1802.00064",
    archivePrefix  = "arXiv",
    primaryClass   = "hep-ph",
    reportNumber   = "DESY-18-017",
    doi            = "10.1140/epjc/s10052-018-6090-8",
    journal        = "Eur. Phys. J. C",
    volume         = "78",
    number         = "8",
    pages          = "621",
    year           = "2018"
}

@article{Bonvini:2018xvt,
    author         = "Bonvini, Marco and Marzani, Simone",
    title          = "{Four-loop splitting functions at small $x$}",
    eprint         = "1805.06460",
    archivePrefix  = "arXiv",
    primaryClass   = "hep-ph",
    doi            = "10.1007/JHEP06(2018)145",
    journal        = "JHEP",
    volume         = "06",
    pages          = "145",
    year           = "2018"
}

@article{MSHT20,
    author = {Bailey, S. and others},
    title = {Parton distributions from {LHC}, {HERA}, {Tevatron} and fixed target
             data: {MSHT20} {PDFs}},
    journal = {Eur. Phys. J. C},
    volume = {81},
    pages = {341},
    year = {2021},
    doi = {10.1140/epjc/s10052-021-09057-0},
    eprint = {2012.04684},
    archivePrefix = {arXiv},
    primaryClass = {hep-ph}
}

@article{Reader2024,
    author = {Reader, Matthew},
    title = {A study of systematic uncertainties within the {MSHT} {PDF} framework},
    journal = {PoS},
    volume = {DIS2024},
    pages = {059},
    year = {2025},
    doi = {10.22323/1.469.0059},
    eprint = {2408.12922},
    archivePrefix = {arXiv},
    primaryClass = {hep-ph}
}

@misc{Barontini:2025,
    author = {Barontini, Andrea and Costantini, Mark N. and
              De Crescenzo, Giovanni and Forte, Stefano and Ubiali, Maria},
    title = {Evaluating the faithfulness of {PDF} uncertainties in the presence
             of inconsistent data},
    eprint = {2503.17447},
    archivePrefix = {arXiv},
    primaryClass = {hep-ph},
    year = {2025},
    note = {arXiv:2503.17447 [hep-ph]}
}

@article{Watt:2012tq,
    author        = {Watt, G. and Thorne, R. S.},
    title         = {Study of {Monte Carlo} approach to experimental uncertainty
                     propagation with {MSTW} 2008 {PDFs}},
    eprint        = {1205.4024},
    archivePrefix = {arXiv},
    primaryClass  = {hep-ph},
    journal       = {JHEP},
    volume        = {08},
    pages         = {052},
    year          = {2012},
    doi           = {10.1007/JHEP08(2012)052}
}

@article{Bailey:2019yze,
    author        = {Bailey, S. and Harland-Lang, L. A.},
    title         = {Differential top quark pair production at the {LHC}:
                     Challenges for {PDF} fits},
    eprint        = {1909.10541},
    archivePrefix = {arXiv},
    primaryClass  = {hep-ph},
    journal       = {Eur. Phys. J. C},
    volume        = {80},
    pages         = {60},
    year          = {2020},
    doi           = {10.1140/epjc/s10052-020-7633-3}
}

@article{CMS:2019WCharm,
    author        = {{CMS Collaboration}},
    title         = {Measurement of associated production of a {$W$} boson and
                     a charm quark in proton-proton collisions at
                     $\sqrt{s}=13$ {TeV}},
    eprint        = {1811.10021},
    archivePrefix = {arXiv},
    primaryClass  = {hep-ex},
    journal       = {Eur. Phys. J. C},
    volume        = {79},
    pages         = {269},
    year          = {2019},
    doi           = {10.1140/epjc/s10052-019-6752-1}
}

@article{CMS:2022JetsPDF,
    author        = {{CMS Collaboration}},
    title         = {Measurement and {QCD} analysis of double-differential
                     inclusive jet cross sections in proton-proton collisions
                     at $\sqrt{s}=13$ {TeV}},
    eprint        = {2111.10431},
    archivePrefix = {arXiv},
    primaryClass  = {hep-ex},
    journal       = {JHEP},
    volume        = {02},
    pages         = {142},
    year          = {2022},
    doi           = {10.1007/JHEP02(2022)142},
    note          = {Addendum: JHEP 12 (2022) 035,
                     doi:10.1007/JHEP12(2022)035}
}

\appendix

\section{Iterative profiling of NPs}
\label{app:profiling}

As described in Sec.~\ref{sec:profiling_bartlett}, minimisation of
$\chi^2(\boldsymbol{a}, \boldsymbol{b})$ can be made easier by
first replacing the nuisance parameters $\boldsymbol{b}$ with their
profiled values at fixed $\boldsymbol{a}$.  Here we describe
an iterative procedure for finding these values.
For fixed $\boldsymbol{a}$, differentiation of
Eq.~(\ref{eq:eoe_xfitter_chi2}) with respect to $b_s$ gives
\begin{equation}
 \sum_i\frac{\Gamma_{is}}{\sigma_i^2}
 \left(r_i-\sum_p\Gamma_{ip}b_p\right)
 -\frac{(1+2\epsilon_s^2)b_s}
 {1+2\epsilon_s^2b_s^2}=0 ,
 \label{eq:gvm_nuisance_equations}
\end{equation}
where $r_i$, $\sigma_i^2$ and $\Gamma_{is}$ are defined in
Eq.~(\ref{eq:profiling_definitions}).
For $\epsilon_s\neq 0$, these coupled equations are nonlinear. Here we implement
the iterative form of the perturbative procedure derived in
Ref.~\cite{Canonero:2024zlb} in matrix form for use in \textsc{xFitter}.
At iteration $k$, we define an effective constraint variance
\begin{equation}
 \hat{\sigma}^{2\,(k)}_{u_s}
 =
 \frac{1+2\epsilon_s^2\left(b_s^{(k)}\right)^2}
      {1+2\epsilon_s^2},
 \label{eq:effective_constraint_variance}
\end{equation}
and construct the matrices
\begin{equation}
 \begin{aligned}
 A_{sp}^{(k)}&=
 \frac{\delta_{sp}}{\hat{\sigma}^{2\,(k)}_{u_s}}
 +\sum_i\frac{\Gamma_{is}\Gamma_{ip}}{\sigma_i^2},\\
 B_s&=\sum_i\frac{\Gamma_{is}r_i}{\sigma_i^2}.
 \end{aligned}
 \label{eq:iterative_AB}
\end{equation}
The updated NPs are obtained from
\begin{equation}
 b_s^{(k+1)}
 =
 \sum_p\left(A^{(k)}\right)^{-1}_{sp}B_p.
 \label{eq:iterative_nuisance_solution}
\end{equation}
At convergence, Eqs.~(\ref{eq:effective_constraint_variance})--(\ref{eq:iterative_nuisance_solution}) solve
Eq.~(\ref{eq:gvm_nuisance_equations}) and therefore minimise the $\chi^2$
with respect to the NPs at fixed $\boldsymbol{a}$. The procedure is initialised with
$\hat{\sigma}^{2\,(0)}_{u_s}=1$. The first iteration therefore
uses
\begin{equation}
 \begin{aligned}
 A_{sp}^{(0)}&=\delta_{sp}
 +\sum_i\frac{\Gamma_{is}\Gamma_{ip}}{\sigma_i^2},\\
 b_s^{(1)}&=\sum_p\left(A^{(0)}\right)^{-1}_{sp}B_p,
 \end{aligned}
 \label{eq:gaussian_nuisance_solution}
\end{equation}
which is precisely the analytic Gaussian profiling used by \textsc{xFitter} for
quadratic NPs. Subsequent
iterations update the effective constraint widths and incorporate the
non-quadratic GVM terms.

Local convergence of the expansion underlying this procedure is
guaranteed provided
\begin{equation}
 2\epsilon_s^2 b_s^2<1\;.
 \label{eq:profiling_convergence}
\end{equation}
In practice, the iterative solution has been observed to converge even
when this condition is not satisfied. In such cases, however, numerical
convergence should be checked by increasing the number of iterations
and verifying that the fitted results stabilise, particularly in the
presence of large NP shifts, large errors-on-errors, or competing local
minima. If convergence is not achieved, the NPs are minimised numerically.

\subsection{Convergence and numerical validation}
\label{sec:appconvergence}

To demonstrate how the iterative profiling converges in practice,
we consider the seven-source configuration obtained by forward
selection from the \mbox{HERAPDF2.0}, NNLO, RTOPT baseline, as reported in
Table~\ref{tab:forward}. At convergence, further iterations leave the NP
values unchanged, and the $\chi^2$ minimisation equations at fixed
$\boldsymbol{a}$ are satisfied. We therefore compare the profiled fit
with a reference fit in which all seven NPs are minimised numerically. In the
profiled fit, \texttt{sysHZComb1053} and \texttt{proc\_tb21}, which have
large fitted shifts, are still minimised numerically,
while the other five NPs are profiled iteratively. We compare the two fits as
the number of profiling iterations is increased.

\begin{figure*}[t]
  \centering
\begin{minipage}{\objectwidthfraction\textwidth}
\centering
  \includegraphics[width=\linewidth]{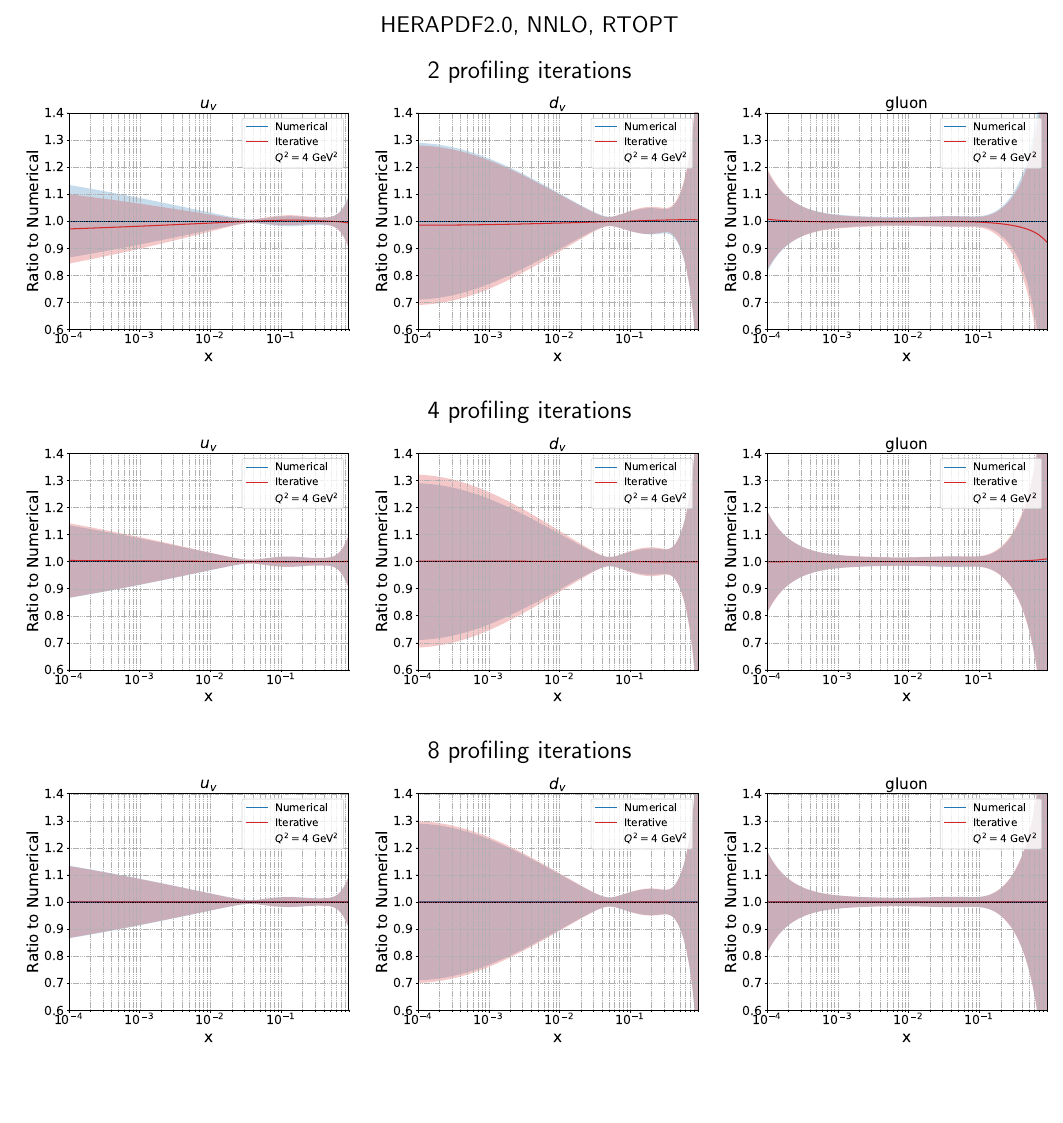}
\caption{Convergence of the iteratively profiled \mbox{HERAPDF2.0}, NNLO, RTOPT fit towards the fit
    in which all seven sources are minimised numerically, at
    $Q^2=4~\mathrm{GeV}^2$.  From left to right, the columns show the ratios
    for $u_v$, $d_v$, and the gluon; from top to bottom, the rows correspond
    to 2, 4, and 8 profiling iterations.}
  \label{fig:convpdfs}
\end{minipage}
\end{figure*}

Figure~\ref{fig:convpdfs} shows this comparison for the $u_v$, $d_v$, and
gluon distributions.  The columns correspond to the three PDFs and the rows
to 2, 4, and 8 profiling iterations.  Each panel shows the ratio of the
profiled fit to the numerically minimized one for the indicated number of
iterations.  Two iterations already give a good approximation to the reference fit;
after four, only a small residual difference remains in the $d_v$
confidence interval, while after eight the profiled and externally
minimised results are indistinguishable at the resolution of the figure.  We therefore use eight
iterations whenever errors-on-errors sources are profiled; the rapid
convergence confirms the numerical accuracy of the iterative procedure.

\setlength{\unitlength}{1.0cm}

\section{Evaluation of the Bartlett correction}
\label{app:bartlett}

The correction terms $b_{\chi^2}$ and $b_{\rm CI}$, entering the
factors defined in Eq.~(\ref{eq:bartlett_correction_coefficients}), follow
from the expansion developed by Lawley~\cite{Lawley:1956} and Cordeiro~\cite{Cordeiro:1993} and
specialised to the GVM in Refs.~\cite{Canonero:2023qiz,Canonero:2024zlb}.
The confidence-interval correction is evaluated for all PDF parameters
together, profiling over the NPs. The resulting factor $C_{\rm CI}$ is the
appropriate one for the joint confidence region of the PDF parameters, and we
use it as a reasonable approximation for the individual confidence intervals,
multiplying their half-widths by $\sqrt{C_{\rm CI}}$. 

The two correction terms are
\begin{equation}
 \begin{aligned}
 b_{\rm CI}&=\sum_s\left(\beta_s^{\rm free}-\beta_s^{\rm fix}\right),\\
 b_{\chi^2}&=\sum_s\left(3\epsilon_s^2-\beta_s^{\rm free}\right),
 \end{aligned}
 \label{eq:bartlett_factors}
\end{equation}
where the sums run over the systematic sources with $\epsilon_s>0$.  Each
source enters through a single quantity $\beta_s$: $\beta_s^{\rm fix}$ is
evaluated with the POIs held fixed at their best-fit values, whereas
$\beta_s^{\rm free}$ is evaluated with them retained as free fit parameters,
so that their correlations with the NPs are included.  The
quantities $\beta_s$ are proportional to $\epsilon_s^2$ and are given by
\begin{equation}
 \begin{aligned}
 \beta_s^{\rm fix}&=\epsilon_s^2
 \left[\,4\,\frac{V_s^{\rm fix}}{\widehat{\sigma}_{u_s}^2}
 -\left(\frac{V_s^{\rm fix}}{\widehat{\sigma}_{u_s}^2}\right)^{\!2}\,\right],\\
 \beta_s^{\rm free}&=\epsilon_s^2
 \left[\,4\,\frac{V_s^{\rm free}}{\widehat{\sigma}_{u_s}^2}
 -\left(\frac{V_s^{\rm free}}{\widehat{\sigma}_{u_s}^2}\right)^{\!2}\,\right].
 \end{aligned}
 \label{eq:bartlett_source_function}
\end{equation}
Here $V_s^{\rm fix}$ and $V_s^{\rm free}$ are the local variances of $\hat{b}_s$, with the PDF
parameters fixed and free respectively.  The common denominator
$\widehat{\sigma}_{u_s}^2$ is the profiled systematic variance in
Eq.~(\ref{eq:gvm_profiled_variance}), evaluated at the fitted shift
$b_s=\widehat{b}_s$.  Each ratio
$V_s/\widehat{\sigma}_{u_s}^2$ is close to unity when the source is constrained
mainly by its auxiliary measurement and decreases as the data determine its
shift more precisely.  Correspondingly, $\beta_s$ ranges from zero to
$3\epsilon_s^2$, which is the maximum contribution from one source.  A source
constrained similarly with the PDFs fixed or free contributes mainly to the
goodness-of-fit correction, whereas an increase in variance when the PDFs are
released transfers the correction to the PDF confidence intervals.  A source
unconstrained by the data in either case contributes to neither.  Since only
the variances enter, a contribution is possible even without a visible shift
of the fitted central values.

The variances $V_s^{\rm fix}$ and $V_s^{\rm free}$ entering
Eq.~(\ref{eq:bartlett_source_function}) are obtained from the covariance
matrix of the fitted parameters, namely the inverse Hessian of the $\chi^2$
at its minimum, evaluated in the quadratic limit with all $\epsilon_s^2$ set
to zero.  In this limit, the Hessian $H$ defined in
Eq.~(\ref{eq:hessian_definition}) has the block form
\begin{equation}
 H=\begin{pmatrix}
 K & G^{\rm T}\\
 G & A
 \end{pmatrix}
 \label{eq:bartlett_block_hessian}
\end{equation}
Here $A$ is the $n_b\times n_b$ NP Hessian at fixed PDFs,
where $n_b$ is the number of fitted NPs. The $n_a\times n_a$ block
$K$ is the PDF-parameter Hessian at fixed NPs, and the $n_b\times n_a$
block $G$ couples the two parameter sets. At the fit minimum,
the three blocks are given by
\begin{equation}
 \begin{aligned}
 A&=\Gamma^{\rm T}W\Gamma+\Lambda,\\
 G&=\Gamma^{\rm T}WD,\\
 K&=D^{\rm T}WD.
 \end{aligned}
 \label{eq:bartlett_hessian_blocks}
\end{equation}
Here $W$ is the inverse covariance matrix of the measurements at the fitted
prediction.  In the absence of statistical covariance blocks it is diagonal,
with entries $1/\sigma_i^2(\widehat{\boldsymbol{a}})$, where $\sigma_i^2$ is
defined in Eq.~(\ref{eq:profiling_definitions}); otherwise, $W$ is the inverse
of the full covariance matrix.
The correlated-shift matrix $\Gamma$, also defined in
Eq.~(\ref{eq:profiling_definitions}), is evaluated at
$\widehat{\boldsymbol{a}}$.  The elements
$D_{im}=\partial T_i/\partial a_m|_{\widehat{\boldsymbol{a}}}$ are evaluated
numerically using finite differences.  Finally, $\Lambda$ is the diagonal
NP constraint matrix, with
$\Lambda_{ss}=1/\widehat{\sigma}_{u_s}^2$.
These expressions neglect the dependence of $\Gamma$ and $W$ on the PDF
parameters, so that they hold in general, independently of the multiplicative
treatment of the uncertainties. We have verified that the neglected terms
affect both the $\chi^2$ and confidence-interval corrections at the
sub-percent level in the examples considered.
Inverting $H$ block by block, the variances entering
Eq.~(\ref{eq:bartlett_source_function}) are
\begin{equation}
 \begin{aligned}
 V_s^{\rm fix}={}&(A^{-1})_{ss},\\
 V_s^{\rm free}={}&\Bigl[A^{-1}+A^{-1}G\\
 &\times\left(K-G^{\rm T}A^{-1}G\right)^{-1}
 G^{\rm T}A^{-1}\Bigr]_{ss}.
 \end{aligned}
 \label{eq:bartlett_variances}
\end{equation}
The second term in $V_s^{\rm free}$ is generated by the mixed block $G$ and
is positive semidefinite, so releasing the PDF parameters can only increase
the NP variance relative to $V_s^{\rm fix}$.  This occurs because a non-zero
$G$ allows a change in the NP to be partly compensated by a correlated change
in the PDFs, leaving the description of the data approximately unchanged.
The two variances therefore coincide when $G=0$, while their difference
quantifies this degeneracy.

\section{NP shifts and uncertainties}
\label{app:hera_np_pulls}

Here we examine the effect of errors-on-errors on NP shifts and their
uncertainties for the seven-source configuration obtained by forward
selection from the \mbox{HERAPDF2.0}, NNLO, RTOPT baseline fit in
Sec.~\ref{sec:robustness} (Table~\ref{tab:forward}), with $\epsilon_s=0.6$
for the selected sources.

The response of these NPs helps explain how the fit accommodates the
data--theory tension. Table~\ref{tab:hera_np_pulls} compares their fitted
shifts and conditional uncertainties before and after introducing
errors-on-errors. The quantities $b_s$ reported here are fitted NP shifts
expressed in units of the nominal systematic uncertainty $\sqrt{v_s}$
in Eq.~(\ref{eq:gvm_full}), with $u_s=0$ and $v_s=1$ in our convention.
Thus, $|b_s|=1$ denotes a displacement equal to one nominal systematic
uncertainty; it does not by itself specify the statistical significance
of that displacement. The conditional uncertainties $\delta b_s$ describe
how precisely the fit determines these shifts. They are expressed in
the same units and evaluated with the PDF parameters held fixed at
each fit minimum.

\begin{inlinetable}
\centering
\captionof{table}{Fitted NP shifts $b_s$ and conditional uncertainties
$\delta b_s$ for the seven NPs selected in the \mbox{HERAPDF2.0}, NNLO,
RTOPT forward-selection fit.}
\label{tab:hera_np_pulls}
\tablebodyfont
\renewcommand{\arraystretch}{1.25}
\setlength{\tabcolsep}{2.5pt}
\begin{tabular*}{\linewidth}{@{\extracolsep{\fill}}lrr@{}}
\toprule
Systematic & \shortstack[r]{Gaussian\\baseline} & Errors-on-errors \\
 & $b_s\pm\delta b_s$ & $b_s\pm\delta b_s$ \\
\midrule
\texttt{sysHZComb1053} & $-2.67 \pm 0.92$ & $-16.41 \pm 2.36$ \\
\texttt{proc\_tb21} & $-3.10 \pm 0.67$ & $-6.31 \pm 1.05$ \\
\texttt{sysHZComb1064} & $+0.71 \pm 0.93$ & $+9.94 \pm 2.76$ \\
\texttt{sysHZComb1114} & $+1.77 \pm 0.81$ & $+4.24 \pm 1.39$ \\
\texttt{sysHZComb1144} & $+1.48 \pm 0.84$ & $+4.08 \pm 1.57$ \\
\texttt{sysHZComb1111} & $-1.84 \pm 0.55$ & $-2.53 \pm 0.67$ \\
\texttt{sysHZComb1113} & $+1.38 \pm 0.84$ & $+3.81 \pm 1.56$ \\
\bottomrule
\end{tabular*}
\end{inlinetable}

Sizeable shifts can indicate sensitivity to errors-on-errors, since the
logarithmic penalty in Eq.~(\ref{eq:gvm_profiled}) grows more slowly than
the Gaussian quadratic penalty at large shifts. One should keep in mind
that, during forward selection, assigning a nonzero $\epsilon_s$ to an
additional systematic source and refitting can change the fit minimum
and modify the shifts of the other NPs. Shifts that are small in the
Gaussian baseline can therefore become sizeable at later steps.
For example, \texttt{sysHZComb1064} is selected even though its
Gaussian-baseline shift is only $0.71$, with a conditional uncertainty
of $0.93$. Conversely, a large shift does not necessarily imply a large
errors-on-errors effect, which also depends on the data constraints
and parameter correlations.

The shift of \texttt{sysHZComb1053} may seem large at about 16 times
the nominal systematic uncertainty. Its size should, however, be
interpreted differently in the GVM than in the Gaussian case.
For $\epsilon_s=0.6$, a deviation of this size under the GVM constraint
is about as rare as a Gaussian fluctuation of 2--3 standard
deviations~\cite{Cowan:2018gqq}.

The conditional uncertainties on the fitted shifts, $\delta b_s$, increase
for all seven selected NPs. In the Gaussian treatment with the PDFs held
fixed, the unit-width auxiliary constraints already limit these
uncertainties to one, and the data can only reduce them further.
This bound does not hold in general in the GVM: the profiled systematic variance
in Eq.~(\ref{eq:gvm_profiled_variance}) increases with $b_s^2$, so the
fitted uncertainties can also grow with the shifts and exceed one.

Another related diagnostic is provided by normalised pulls, which
combine information from the NP shifts and their uncertainties.
In the Gaussian baseline, a source that is weakly constrained by the
data is expected to shift only slightly. A sizeable shift accompanied
by little reduction in uncertainty therefore gives a large pull.
Following the HERA combination definition~\cite{H1:2015ubc},
with $u_s=0$ and $v_s=1$, one can define
\begin{equation}
 p_s=\frac{b_s}{\sqrt{1-(\delta b_s)^2}}.
 \label{eq:gaussian_np_pull}
\end{equation}

These pulls can therefore help identify sources worth testing for
errors-on-errors. However, this definition cannot be applied generally
once errors-on-errors are introduced: the uncertainties $\delta b_s$
can exceed unity, making the argument of the square root negative.
We therefore use the NP shifts rather than the pulls to compare the
Gaussian and
errors-on-errors fits.
\end{document}